\documentclass[prd,twocolumn,superscriptaddress,floatfix,nofootinbib]{revtex4-2}
\pdfoutput=1 % if your are submitting a pdflatex (i.e. if you have
\usepackage[T1]{fontenc} % if needed
\usepackage{amsmath,amssymb,amsfonts,amsthm}
\usepackage{graphicx}
\usepackage[usenames,dvipsnames]{color}
\usepackage{enumitem}
\usepackage{verbatim}
\usepackage[percent]{overpic}
\usepackage{rotating}
\usepackage{hyperref}
\usepackage{array}
\usepackage{mathtools}
\usepackage{lmodern}
\usepackage[normalem]{ulem}
\usepackage{braket}
\usepackage{tensor}
\usepackage{dsfont}
\usepackage{bm}
\usepackage{tikz}
\usepackage{appendix}
\usepackage{accents}

\usepackage{mathrsfs}

\usetikzlibrary{decorations.pathmorphing}
\usepackage{CJKutf8}

\usepackage{soul, xcolor}
\setstcolor{red}

\usetikzlibrary{positioning}
\usetikzlibrary{calc,through,backgrounds}

\theoremstyle{definition}
\newtheorem{definition}{Definition}[section]
\newtheorem{theorem}{Theorem}[section]

\allowdisplaybreaks

\DeclareMathOperator{\arsinh}{arsinh}
\DeclareMathOperator{\arcosh}{arcosh}
\DeclareMathOperator{\artanh}{artanh}

\DeclareMathOperator{\Realpart}{Re}

\begin{document}

\title{Probing Hidden Topology with Quantum Detectors}

\author{Dyuman Bhattacharya}
\email{d7bhatta@uwaterloo.ca}
\affiliation{Department of Physics and Astronomy, University of Waterloo, Waterloo, Ontario, N2L 3G1, Canada}

\author{Jorma Louko}
\email{jorma.louko@nottingham.ac.uk} 

\affiliation{School of Mathematical Sciences, University of Nottingham, Nottingham NG7 2RD, United Kingdom}

\author{Robert B. Mann}
\email{rbmann@uwaterloo.ca}
\affiliation{Department of Physics and Astronomy, University of Waterloo, Waterloo, Ontario, N2L 3G1, Canada}
\affiliation{Institute for Quantum Computing, University of Waterloo, Waterloo, Ontario, N2L 3G1, Canada}
\affiliation{Perimeter Institute for Theoretical Physics,  Waterloo, Ontario, N2L 2Y5, Canada}

\date{October 2024. Updated February 2025.\\ aaPublished in Phys.\ Rev.\ D \textbf{111}, 045005 (2025), doi.org/10.1103/PhysRevD.111.045005.\\ aaFor Open Access purposes, this Author Accepted Manuscript is made available under CC BY public copyright.}

\begin{abstract}
We consider the transition rate of a static Unruh-DeWitt detector in two $(2+1)$-dimensional 
black hole spacetimes that are 
isometric to the static Ba\~nados-Teitelboim-Zanelli 
black hole outside the horizon but have no asymptotically locally anti-de Sitter exterior behind the horizon. 
The spacetimes are the $\mathbb{R}\text{P}^{2}$ geon, with spatial topology $\mathbb{R}\text{P}^{2}\setminus\{\text{point at infinity}\}$, 
and the Swedish geon of \AA{}minneborg \emph{et al\/}, 
with spatial topology $T^{2}\setminus\{\text{point at infinity}\}$. 
For a conformal scalar field, prepared in the Hartle-Hawking-type state that is induced from the global vacuum on the anti-de Sitter covering space, we show numerically that the detector's transition rate distinguishes the two spacetimes, particularly at late exterior times, and we trace this phenomenon to the differences in the isometries that are broken by the quotient construction from the universal covering space. Our results provide an example in which information about the interior topology of a black hole is accessible to a quantum observer outside the black hole. 
\end{abstract}

\maketitle
\flushbottom

\section{Introduction}

Black holes are amongst the most fascinating objects in nature, in large part because they cloak certain regions of space and time from all distant observers. 
While black holes created in a gravitational collapse are not time symmetric, their eternal counterparts contain both a future horizon and past horizon, and continuing the spacetime analytically past these horizons reveals new structure, a prime example of which is the second asymptotically flat region in the Kruskal(-Szekeres-Fronsdal) \cite{PhysRev.119.1743,szekeres-reprint,PhysRev.116.778}
extension of positive mass Schwarzschild spacetime.  
The structure past the horizon has a pivotal role in recognising Hawking temperature as the distinguished temperature for quantum fields in thermal equilibrium with the black hole~\cite{Hartle:1976tp,Israel:1976ur}, 
and it provides the cornerstone to the view of black hole entropy as a fundamentally quantum-gravitational concept~\cite{Gibbons:1976ue}. 

What lies beyond the black hole horizon is, by definition, inaccessible to classical observers outside the horizon. However, in an eternal black hole spacetime, with both a future horizon and a past horizon, states of quantum fields that are regular on the extended spacetime are constrained by the spacetime structure behind the horizons. 
For such states, it is therefore possible for a local quantum observer to distinguish what lies behind the horizon by probing the state of a quantum field outside the horizon \cite{Louko:1998dj,Louko:1998hc,Louko:2000tp,Bruschi:2010rw,Louko:2010tq,Smith_2014}. In this paper we address this phenomenon for a class of black hole spacetimes known as topological geons~\cite{Sorkin1986}, 
in which there is no second asymptotic region behind the horizon \cite{Louko:2010tq,Louko_2005,Kottanattu:2010zg}.

The geon black hole spacetimes that we consider are $(2+1)$-dimensional eternal black holes, which are solutions to Einstein's equations with a negative cosmological constant, and have an exterior region that is isometric to that of the spinless non-extremal Ba\~nados-Teitelboim-Zanelli (BTZ) black hole \cite{PhysRevLett.69.1849,Banados:1992gq}.We consider two families, both obtained as a quotient of an open set in $(2+1)$-dimensional anti-de~Sitter spacetime (AdS${}_3$) by a discrete group of isometries. 
One family is the $\mathbb{R}\text{P}^{2}$ geon~\cite{Louko:1998hc}, which is a 
$\mathbb{Z}_2$ quotient of the BTZ black hole and has spatial topology $\mathbb{R}\text{P}^{2}\setminus\{\text{point at infinity}\}$. 
The other is the ``Swedish geon'' of \AA{}minneborg \emph{et al\/}~\cite{aaminneborg_1998}, 
which has spatial topology $T^{2}\setminus\{\text{point at infinity}\}$. In both of these spacetimes, the exterior Killing vector that generates exterior BTZ time translations does not extend to a Killing vector on the full spacetime. 
The state that a quantum field inherits from the global vacuum on AdS${}_3$ is hence not static in the exterior region~\cite{Louko:1998hc,Louko:2000tp}, 
in contrast to the Hartle-Hawking-type state inherited by the BTZ spacetime \cite{LifschytzBTZ,Carlip:1995qv}, 
and this nonstaticity bears an imprint of the spacetime structure behind the horizon. We shall analyse how a local quantum observer outside the horizon can access this imprint. 

We consider a massless conformal scalar field. We model a local quantum observer by an Unruh-DeWitt (UDW) detector \cite{Unruh1979evaporation, DeWitt1979}, a pointlike two-level system that is linearly coupled to the field, and we work to linear order in the coupling. We compute the detector's transition rate on a trajectory that is static in the exterior geometry: the time dependence in this transition rate is then precisely due to the geon-like spacetime structure behind the horizon. 

For both the $\mathbb{R}\text{P}^{2}$ geon and the Swedish geon, our numerical results show that the transition rate in the asymptotic past reduces to that on the BTZ hole. Assuming the detector to be switched on in the asymptotic past, the late time behaviour, however, differs for the two spacetimes. The late time transition rate in the Swedish geon asymptotes to that in the BTZ hole. The late time transition rate in the $\mathbb{R}\text{P}^{2}$ geon, by contrast, asymptotes to a different constant for a gapless detector, and it displays persisting oscillations for a detector with a nonvanishing gap; the latter property is consistent with what was previously found for the $\mathbb{R}\text{P}^{2}$ geon~\cite{Smith_2014}, 
and with what happens for uniformly accelerated detectors in flat spacetimes with a similar structure behind the Rindler horizon \cite{Langlois:2005nf,Langlois:2005if}. 
This difference between the late time transition rates is a dramatic demonstration of how the detector in the exterior can distinguish different spacetime topologies behind the horizon. 

As a side product, we give explicit formulas for the isometries by which the Swedish geon is quotiented from $\text{AdS}_{3}$, in a representation in which the exterior region takes a standard form, and we display a fundamental domain for the spacelike hypersurface of time symmetry. This representation of the isometries can be employed to investigate other properties of the quantum state that the Swedish geon inherits from the global vacuum on $\text{AdS}_{3}$, such as the stress-energy tensor. 

Our paper is organized as follows. Section \ref{sec:udw-detector} gives a brief review of an UDW detector coupled to a scalar field, writing down the transition rate of a detector that is switched on and off sharply. 
Section \ref{sec:quotients} presents the static BTZ black hole, the $\mathbb{R}\text{P}^{2}$ geon and the Swedish geon as quotients of subsets of $\text{AdS}_3$, deferring technicalities to three appendices. Section \ref{sec:response-on-quotients} gives the image sum constructions for the static detector's transition rate on these spacetimes, for a conformal scalar field in the state induced from the global vacuum on $\text{AdS}_3$. Numerical results for the detector's transition rate are presented in Section~\ref{sec:results}. Section \ref{sec:conclusions} gives the conclusions and a brief discussion.

\section{Unruh-DeWitt detector}
\label{sec:udw-detector}

We probe the quantum field by an Unruh-DeWitt (UDW) detector, which is a point-like quantum system that has two states: a state $\ket{0}_{D}$ with energy $0$, and a state $\ket{1}_D$ with energy~$\Omega$, where $\Omega$ is a real-valued parameter. 
For $\Omega>0$, $\ket{0}_{D}$ is the ground state and $\ket{1}_D$ is the excited state; 
for $\Omega<0$, the roles of $\ket{0}_{D}$ and $\ket{1}_{D}$ are reversed. 

The detector follows a spacetime trajectory $x_{D}(\tau)$, which we parametrize by its proper time $\tau$, and it is coupled to a real scalar field $\phi(x)$ through the interaction Hamiltonian
\begin{align}
    H_{I}(\tau)=\lambda\chi(\tau)\mu(\tau)\otimes\phi[x_D(\tau)],
\end{align}
where $\lambda$ is a coupling constant, $\chi(\tau)$ is a switching function that determines how the interaction is turned on and off, and $\mu(\tau)$ is the detector's monopole moment operator, given by 
\begin{align}
\mu(\tau)=e^{i\Omega\tau}\sigma^{+}+e^{-i\Omega\tau}\sigma^{-} \,,
\end{align}
where $\sigma^{+}=\ket{1}_{D}\bra{0}_D$ and $\sigma^{-}=\ket{0}_{D}\bra{1}_D$ are the respective raising and lowering ladder operators. 
The detector's Hilbert space is $\mathbb{C}^2$. 

Before the interaction is turned on, we prepare the detector in the state $\ket{0}_{D}$ and the field $\phi$ in the state that we denote here by $\ket{0}$ and specify individually in the applications in the later sections. 
Working in first-order perturbation theory, the probability to find the detector in the state $\ket{1}_{D}$ after the interaction has ceased, regardless the final state of the field, is \cite{Junker:2001gx,Louko:2007mu}
\begin{align}
P(\Omega) = \lambda^{2}
\mathcal{F}(\Omega), 
\end{align}
where the response function $\mathcal{F}(\Omega)$ is given by 
\begin{align}
\mathcal{F}(\Omega) = \int d\tau& d\tau'\chi(\tau)\chi(\tau')
\, e^{-i\Omega(\tau-\tau')}W(\tau,\tau'),
\end{align}
and 
$W(\tau,\tau')= \bra{0} \phi \bigl(x_{D}(\tau) \bigr),\phi\bigl(x_D(\tau')\bigr) \ket{0}$ is the pull-back of the field's Wightman function to the detector's worldline. 

We take $\chi$ to be the characteristic function of an interval of time, as the response function remains well defined in this limit in $2+1$ spacetime dimensions despite the discontinuities at the switch-on and switch-off moments. We further consider not the response function itself but its derivative with respect to the switch-off moment, denoted by~$\dot{\mathcal{F}}$, which has an interpretation as (a multiple of) the detector's transition rate, operationally measurable using ensembles of detectors~\cite{Langlois:2005if,Satz:2006kb,Louko:2007mu}. 
For a detector switched on at time $\tau_0$ and off at time~$\tau$, 
the transition rate is given by \cite{Hodgkinson:2012mr}
\begin{align}
    \dot{\mathcal{F}}(\Omega)=\frac{1}{4}+2\int_{0}^{\Delta\tau}ds \Realpart\big[e^{-i\Omega s}W(\tau,\tau-s)\big], 
\label{eq:transition-rate}
\end{align}
where $\Delta\tau=\tau-\tau_{0}$. 
If the detector is switched on in the asymptotic past, $\tau_0\rightarrow-\infty$, 
the transition rate \eqref{eq:transition-rate} becomes 
\begin{align}
    \dot{\mathcal{F}}(\Omega)=\frac{1}{4}+2\int_{0}^{\infty}ds \Realpart\big[e^{-i\Omega s}W(\tau,\tau-s)\big].
\label{eq:transitionrate}
\end{align}

Note that while $W(\tau,\tau')$ has a distributional singularity at $\tau=\tau'$, the integrals in \eqref{eq:transition-rate} and \eqref{eq:transitionrate} do not have distributional contributions from $s=0$. 
This is because 
the singularity in $W(\tau,\tau')$ is that of 
\begin{align}
\frac{-i} {4\pi(\tau-\tau' - i \epsilon)} \xrightarrow[\epsilon\to0_+]{} 
\frac{-i} {4\pi} P \! \left(\frac{1}{\tau-\tau'}\right) 
+ \frac{1}{4}\delta(\tau-\tau'), 
\label{eq:inertial-minkowski-wightman}
\end{align}
where $P$ stands for the Cauchy principal value: the term $\frac14$ in \eqref{eq:transition-rate} and \eqref{eq:transitionrate} has come from the Dirac delta in \eqref{eq:inertial-minkowski-wightman}, and the integrals in \eqref{eq:transition-rate} and \eqref{eq:transitionrate} are nonsingular at $s=0$ on taking the real part.

\section{Quotients of $\text{AdS}_3$}
\label{sec:quotients}

In this section we describe three black hole spacetimes that are quotients of an open subset in $\text{AdS}_3$. 

\subsection{$\text{AdS}_3$}

Recall that $\text{AdS}_{3}$ may be defined as the hyperboloid 
\begin{align}
    X_{1}^2+X_{2}^2-T_{1}^2-T_{2}^2=-\ell^2 
\label{eq:embedding-hyperboloid}
\end{align}
in $\mathbb{R}^{2,2}$ with the metric 
\begin{align}
    ds^2=dX_{1}^2+dX_{2}^2-dT_{1}^{2}-dT_{2}^2, 
\label{eq:2+2embeddingmetric}
\end{align}
where $\ell$ is a positive constant of dimension length. 
This spacetime is a solution to Einstein's equations with the cosmological constant $\Lambda=-\ell^2$ \cite{Carlip:1995qv}. 

It is useful to introduce two local coordinate systems on open subsets of $\text{AdS}_{3}$ as follows. 

First, in the subset $\textsf{B}_1$, defined as $T_1 > |X_1|$, we introduce the coordinates $(U,V,\phi)$ by 
\begin{align}
\begin{split}
T_1 
&= 
\ell 
\left(
\frac{1-UV}{1+UV}
\right)
\cosh(\sqrt{M}\,\phi), 
\\
X_1
&=
\ell 
\left(
\frac{1-UV}{1+UV}
\right)
\sinh(\sqrt{M}\,\phi), 
\\
T_2
&=
\ell
\left(
\frac{V+U}{1+UV}
\right), 
\\
X_2
&=
\ell
\left(
\frac{V-U}{1+UV}
\right), 
\end{split}
\end{align}
where $-1 < UV < 1$, $-\infty < \phi < \infty$, and $M$ is a positive parameter. 
The metric takes the form 
\begin{align}
    ds^2=-\frac{4\ell^2}{(1+UV)^2}dUdV+r_{h}^2\bigg(\frac{1-UV}{1+UV}\bigg)^2d\phi^2.
\label{eq:kruskalmetric}
\end{align}

Second, in the subset of $\textsf{B}_1$ where $U<0$ and $V>0$, we introduce the coordinates $(t,r,\phi)$, where 
\begin{align}
    \begin{split}
        \frac{r}{r_{h}}&=\frac{1-UV}{1+UV},\\
        \frac{r_{h}t}{\ell^2}&= \frac12 \ln \! \left(-\frac{V}{U}\right), 
    \end{split}
\end{align}
where $r_h = \ell\sqrt{M}$, so that $r_h < r < \infty$ and $-\infty < \phi < \infty$. In terms of the embedding space coordinates, this means 
\begin{align}
    \begin{split}
        X_1&=\ell\frac{r}{r_h}\sinh \! \bigg(\frac{r_h}{\ell}\phi\bigg),\\
        X_2&=\ell\sqrt{\frac{r^2}{r_h^2}-1}\cosh \! \bigg(\frac{r_h}{\ell^2}t\bigg),\\
        T_1&=\ell\frac{r}{r_h}\cosh \! \bigg(\frac{r_h}{\ell}\phi\bigg),\\
        T_2&=\ell\sqrt{\frac{r^2}{r_h^2}-1}\sinh \! \bigg(\frac{r_h}{\ell^2}t\bigg),\\
    \end{split}
\label{eq:BTZcoords-embedding}
\end{align}
and the metric reads 
\begin{align}
    ds^{2}=-\bigg(\frac{r^2-r_{h}^2}{l^2}\bigg)dt^2+\bigg(\frac{r^2-r_{h}^2}{l^2}\bigg)^{-1}dr^2+r^2d\phi^2. 
\label{eq:BTZmetric}
\end{align}
We refer to \eqref{eq:BTZmetric} as the AdS-Rindler metric~\cite{Smith.2017}. 

$\phantom{xxx}$ % to get the vertical spacing right 

\subsection{BTZ black hole}
\label{subsec:BTZquotient}

The spinless BTZ spacetime is obtained from $\textsf{B}_1$ with the metric \eqref{eq:kruskalmetric} by the identification $(U,V,\phi) \sim (U,V,\phi+2\pi)$. 
This spacetime is an eternal black-and-white-hole spacetime, 
with a bifurcate Killing horizon at $|U|=|V|$, and $(U,V,\phi)$ are Kruskal-type coordinates \cite{PhysRevLett.69.1849,Banados:1992gq,Carlip:1995qv}. The mass of the black hole is~$M$. 

We may write the spacetime as the quotient space 
\begin{align}
\mathcal{M}_{\text{BTZ}}=\textsf{B}_1/\mathbb{Z}, 
\end{align}
where $\mathbb{Z}\simeq\{\Gamma^{n}: n\in\mathbb{Z}\}$ with 
\begin{align}
    \Gamma:\;(U,V,\phi)\mapsto(U,V,\phi+2\pi). 
\label{eq:Gamma-action}
\end{align}

The quadrant where $U<0$ and $V>0$, covered by the coordinates $(t,r,\phi)$ with the identification $(t,r,\phi)\sim(t,r,\phi+2\pi)$, is an exterior region. The metric is given by~\eqref{eq:BTZmetric}, and the horizon is at $r\to r_h$.

\subsection{$\mathbb{R}\text{P}^{2}$ geon}
\label{subsec:rp2geonquotient}

The $\mathbb{R}\text{P}^{2}$ geon \cite{Louko:1998hc} is the quotient of the BTZ spacetime by the isometry group 
$\{e,J\}\simeq\mathbb{Z}_2$, where $e$ is the identity and 
\begin{align}
    J: (U,V,\phi)\mapsto(V,U,\phi+\pi) . 
\label{eq:J-action}
\end{align}
We may write 
\begin{align}
    \mathcal{M}_{\mathbb{R}\text{P}^{2}}=\mathcal{M}_{\text{BTZ}}/\mathbb{Z}_2.
\end{align}
This spacetime is an eternal black and white hole spacetime with just a single exterior region, in which the metric is as in~\eqref{eq:BTZmetric}. While the exterior is static, with the timelike Killing vector $\partial_t$, this Killing vector does not extend to all of the geon spacetime, and the geometry behind the horizons hence makes the $t=0$ hypersurface a distinguished hypersurface of time symmetry \cite{Louko:1998hc,Smith_2014,Smith.2017}.

\subsection{Swedish geon}
\label{subsec:swedishgeonquotient}

The Swedish geon \cite{aaminneborg_1998} is a single-exterior black-and-white hole
quotient of 
an open region in $\text{AdS}_3$, with the exterior metric given by the BTZ exterior~\eqref{eq:BTZmetric}, but with behind-the-horizon identifications that give the spacetime the spatial topology of a punctured torus, the puncture being at the AdS infinity, $r\to\infty$ in~\eqref{eq:BTZmetric}. 
We review the construction of this spacetime in the three appendices, showing that the spacetime is a quotient of a subset of $\text{AdS}_3$ by the free group generated by the $O(2,2)$ matrices $A$ and $B$ given by 
\begin{widetext}
\begin{subequations}
\label{eq:A-B-maintext}
\begin{align}
        A&=
        \begin{pmatrix}
            1&0&0&0\\[1ex]
            0&{\displaystyle\frac{\cosh(6z)+2\cosh(4z)+\cosh(2z)}{4 \sinh^2(z)}}&{\displaystyle\frac{\sqrt{\cosh(2z)}\cosh(4z)}{\sinh^2(z)}}&\sinh(4z)\\[3ex]
            0&{\displaystyle-\frac{\sqrt{\cosh(2z)}}{\sinh^2(z)}}&- \coth^2(z)&0\\[3ex]
            0&{\displaystyle-\frac{\cosh(5z)+3\cosh(3z)+4\cosh(z)}{2\sinh(z)}}&-4[\cosh(2z)]^{3/2}\coth(z)&-\cosh(4z)
        \end{pmatrix},\\[2ex]
        B&=
        \begin{pmatrix}
            1&0&0&0\\[1ex]
            0&{\displaystyle\frac{\cosh(4z)+\cosh(2z)}{2 \sinh^2(z)}}&{\displaystyle\frac{[\cosh(2z)]^{3/2}}{\sinh^2(z)}}&{\displaystyle\frac{\cosh(3z)+\cosh(z)}{\sinh(z)}}\\[3ex]
            0&{\displaystyle-\frac{[\cosh(2z)]^{3/2}}{\sinh^2(z)}}&- \coth^2(z)&-2\sqrt{\cosh(2z)}\coth(z)\\[3ex]
            0&{\displaystyle\frac{\cosh(3z)+\cosh(z)}{\sinh(z)}}&2\sqrt{\cosh(2z)}\coth(z)
            &2\cosh(2z)+1
        \end{pmatrix},
\end{align}
\end{subequations}
\end{widetext}
acting by matrix multiplication on the column vectors 
\begin{align}
    \begin{pmatrix}
        T_2\\
        T_1\\
        X_2\\
        X_1
        \end{pmatrix} , 
\end{align}
where $z=\frac14 \pi\sqrt{M}$. 
(We have dropped in \eqref{eq:A-B-maintext} the undertildes by which these matrices are denoted in \eqref{eq:Atilde-Btilde} in Appendix \ref{sec:generators}.) 
The exterior region is again covered by the coordinates $(t,r,\phi)$ \eqref{eq:BTZcoords-embedding}, so that $M$ is the black hole mass. The isometry $(t,r,\phi) \mapsto (t,r,\phi +2\pi)$ is generated by the matrix $B A B^{-1} A^{-1}$. 

\section{Transition rate comparisons} 
\label{sec:response-on-quotients}

In this section we present the image sum expressions for the detector's transition rate on the BTZ black hole, the $\mathbb{R}\text{P}^{2}$ geon and the Swedish 
geon, in the state induced from the $\text{AdS}_3$ global vacuum. 

% \vspace{4ex}

\subsection{Preamble: choice of the state}

To begin, recall that Fock quantisation of a real Klein-Gordon field relies on a decomposition of the complexified classical solution space into a positive Klein-Gordon norm subspace and a negative Klein-Gordon norm subspace. In Minkowski spacetime, choosing the positive norm subspace to be the positive frequency subspace with respect to Minkowski time translations results in the standard Minkowski Fock vacuum as the ground state of the theory, and this vacuum is physically interpreted as the no-particle state for all inertial observers~\cite{birrell}. 

In curved spacetime, by contrast, there is no general criterion for choosing the positive norm subspace. In curved spacetimes with symmetries or asymptotic symmetries, however, the symmetries may offer a criterion for this choice, reflecting a choice of physical input. An example is de Sitter space, where de Sitter invariance combined with technical regularity properties singles out the Fock vacuum known as the Euclidean, Chernikov-Tagirov or Bunch-Davies vacuum \cite{AllenVacuumStatesdS,ChernikovTagirov,BunchDavies}, which has a physical interpretation as a thermal state for inertial observers~\cite{GibbonsHawkingDeSitter}. 
Another example is an expanding cosmology with an early time asymptotically de Sitter era, where the asymptotic early time symmetries single out a Bunch-Davies state~\cite{Mukhanov:2007zz}.

An example closer to the spacetimes considered in the present paper is the Schwarzschild black hole spacetime. 
In the exterior region, Schwarzschild time translations single out the state known as the Boulware state~\cite{Boulware:1974dm}, which is close to the Minkowski vacuum at asymptotically flat infinity, but is singular on approaching the black and white hole horizons. 
On the Kruskal extension of Schwarzschild, translations along the horizon generator affine parameters single out the state known as the Hartle-Hawking or Hartle-Hawking-Israel state~\cite{Hartle:1976tp,Israel:1976ur}, which describes physically a Hawking-radiating black hole in thermal equilibrium with a heat bath at infinity, 
and is the unique regular state that is invariant under all the continuous isometries~\cite{Kay:1988mu}.
A third state of interest is the Unruh state~\cite{Unruh1979evaporation}, 
defined on the union of an exterior region, the black hole region, and the black hole horizon joining the two: this state is defined in terms of positive frequency with respect to Minkowski time translations at the past infinity and with respect to the future horizon generator on the future horizon, and it describes physically a Hawking-radiating black hole with no infalling radiation from the past infinity, but it is singular on the past horizon. 

On the spinless BTZ black hole spacetime, a new feature compared with Schwarzschild is that the asymptotically AdS infinity requires a boundary condition to make the theory unitary~\cite{Avis:1977yn}. Physically, the effect of   asymptotically AdS infinity on wave propagation is similar to that of a boundary at a finite distance, and the boundary condition means specifying the phase shift for waves that are reflected back from  infinity. The boundary condition therefore does not allow the field to be decomposed into ``incoming'' and ``outgoing'' parts near  infinity, and this prevents the construction of a time-asymmetric state similar to the Unruh state on Schwarzschild. A~Boulware-type state in the BTZ black hole exterior can be readily defined, using the boundary condition at infinity and the notion of positive frequency with respect to the exterior Killing time translations, but this state is singular on the future and past horizons, like the Boulware state on exterior  Schwarzschild. 

The state we consider on the spinless BTZ black hole spacetime is the state induced from the $\text{AdS}_3$ global vacuum by the $\mathbb{Z}$ quotient described in Section \ref{subsec:BTZquotient} \cite{Steif:1993zv}. We may refer to this state as a Hartle-Hawking type state: it is regular over all of the BTZ spacetime, it is invariant under the continuous isometries of the spacetime, and it describes physically a Hawking-radiating black hole in thermal equilibrium with a heat bath at   asymptotic infinity. 
We emphasise that this state knows not just about the geometry in a single exterior region: it knows about the full global structure that the spacetime has behind the black and white hole horizons. Detectors interacting with this state have been considered in \cite{Hodgkinson:2012mr,Preciado-Rivas:2024gzm}. 

On the $\mathbb{R}\text{P}^{2}$ geon and the Swedish geon, we follow the example of the BTZ spacetime and consider the state induced from the $\text{AdS}_3$ global vacuum by the quotients described in Sections \ref{subsec:rp2geonquotient} and~\ref{subsec:swedishgeonquotient}. 
In each case, the state is regular on all of the spacetime, including the horizons, and the state knows about the full global spacetime structure, including that behind the horizons. 
What is new is that the restriction of the state to the exterior region is not invariant under the exterior Killing time translations, because the quotienting group contains elements that do not commute with these translations. 
Detectors interacting with this state on the $\mathbb{R}\text{P}^{2}$ geon have been considered in \cite{Smith_2014,Smith.2017,Spadafora:2024gqj}. We shall consider a detector interacting with this state on the Swedish geon, and compare with detectors on the BTZ spacetime and on the $\mathbb{R}\text{P}^{2}$ geon.

\subsection{Image sum in the state induced from the $\text{AdS}_3$ global vacuum}

On each of the three quotients of an open region in $\text{AdS}_3$, we now consider the state induced from the $\text{AdS}_3$ global vacuum, as described above. 

We denote by $G$ the discrete group by which the quotient is taken. The Wightman function on the quotient spacetime is then given by the method-of-images sum 
\begin{equation}
    W_{\text{AdS}_3/G}(x, x^{\prime})=\sum_{g\in G} W_{\text{AdS}_3}(x, g x^{\prime}), 
\label{eq:wightman-imagesum}
\end{equation}
where $W_{\text{AdS}_3}$ is the Wightman function on $\text{AdS}_3$. 

The global vacuum Wightman function for a massless conformally-coupled scalar field in $\mathrm{AdS_3}$ takes the form \cite{Carlip:1995qv,Smith.2017} 
\begin{equation}
    W_{\mathrm{AdS_3}}(x,x')=\frac{1}{4\pi\sqrt{2}\ell}\Bigg(\frac{1}{\sqrt{\sigma(x,x')}}-\frac{\zeta}{\sqrt{\sigma(x,x')+2}}\Bigg) , 
\label{eq:AdS-wightman}
\end{equation}
where $\sigma(x,x')$ is $1/(2\ell^2)$ times the square of the geodesic distance between the spacetime points $x$ and $x'$ in the embedding space $\mathbb{R}^{2,2}$, 
\begin{align}
    \begin{split}
        \sigma(x,x')=\frac{1}{2\ell^2}\Big[&(X_1-X_1')^2-(T_1-T_1')^2\\&+(X_2-X_2')^2-(T_2-T_2')^2\Big] , 
    \end{split}
\end{align}
and the parameter $\zeta\in\{ -1, 0, 1 \}$ specifies the boundary condition at the asymptotically AdS infinity. $\zeta=1$ and $\zeta=-1$ are respectively the Dirichlet and Neumann boundary conditions, each of which determines a quantum theory with unitary evolution. $\zeta=0$ is the transparent boundary condition, with no reflection from the infinity, and a less clear physical interpretation~\cite{Avis:1977yn}. 

In all three spacetimes, 
we consider a UDW detector on the static exterior worldline that is given in the BTZ exterior coordinates \eqref{eq:BTZmetric} by 
\begin{align}
    (t,r,\phi) =\bigg(\frac{\ell}{\sqrt{R^{2}-r_{h}^{2}}}\tau,\; R,\; 0\bigg) , 
\label{eq:static-trajectory}
\end{align}
where $R>r_h$ is the detector's radial coordinate and $\tau$ is the detector's proper time. Taking the detector to be switched on in the asymptotic past, the detector's transition rate may then be obtained by inserting the trajectory \eqref{eq:static-trajectory} in \eqref{eq:transitionrate} with \eqref{eq:wightman-imagesum} and~\eqref{eq:AdS-wightman}. 

We consider the three spacetimes in turn. 

\subsection{BTZ}

On the BTZ spacetime, we have 
\begin{align}
    W_{\text{BTZ}}(x,x')=\sum_{n=-\infty}^{\infty}W_{\text{AdS}_{3}}(x,\Gamma^{n}x'),
\end{align}
where $\Gamma$ is given in~\eqref{eq:Gamma-action}. In the exterior, in the coordinates \eqref{eq:BTZcoords-embedding}, the action of $\Gamma$ reads 
\begin{align}
    \Gamma:\;(t,r,\phi)\mapsto(t,r,\phi+2\pi). 
\label{eq:Gamma-exterior-action}
\end{align}
It follows that \cite{Hodgkinson:2012mr}
\begin{align}
    \begin{split}
        \dot{\mathcal{F}}_{\text{BTZ}}(\Omega)=&\frac{e^{-\ell\beta\Omega/2}}{2\pi}\sum_{n=-\infty}^{\infty}\int_{0}^{\infty}dy\cos(y\ell\beta\Omega/2)\\
        &\times\bigg(\frac{1}{\sqrt{K_{n}+\cosh^{2} \! y}}-\frac{\zeta}{\sqrt{Q_{n}+\cosh^{2} \! y}}\bigg),
    \end{split}
\end{align}
where
\begin{align}
    \begin{split}
        K_{n}&:=\frac{R^{2}}{R^{2}-r_{h}^2}\sinh^{2} \! \bigg(n\pi\frac{r_{h}}{\ell}\bigg),\\
        Q_{n}&:=K_{n}+\frac{r_{h}^2}{R^{2}-r_{h}^{2}},\\
        \beta&:=2\pi\frac{\sqrt{R^{2}-r_{h}^{2}}}{r_h}.
    \end{split}
\end{align}
Note that the transition rate has no dependence on the switch-off moment~$\tau$. 
This is because the isometry $\Gamma$ \eqref{eq:Gamma-action} commutes with translations along the trajectory~\eqref{eq:static-trajectory}.

\subsection{$\mathbb{R}\text{P}^{2}$ geon}

On the $\mathbb{R}\text{P}^{2}$ geon, we have 
\begin{align}
    W_{\mathbb{R}\text{P}^{2}}(x,x')=W_{\text{BTZ}}(x,x')+W_{\text{BTZ}}(x,Jx'), 
\end{align}
where $J$ is given in~\eqref{eq:J-action}. 
It follows that 
\begin{align}
    \dot{\mathcal{F}}_{\mathbb{R}\text{P}^{2}}(\Omega)=\dot{\mathcal{F}}_{\text{BTZ}}(\Omega)+\Delta\dot{\mathcal{F}}_{\mathbb{R}\text{P}^{2}}(\Omega,\tau),
\end{align}
where the additional contribution due to the geon identification may be written as \cite{Smith_2014}
\begin{align}
    \begin{split}
        \Delta\dot{\mathcal{F}}_{\mathbb{R}\text{P}^{2}}(\Omega,\tau)&=\frac{1}{2\pi}\sum_{n=-\infty}^{\infty}\int_{0}^{\infty}dy\cos(y\ell\beta\Omega/2)\\
        &\times\Bigg[\frac{1}{\sqrt{\bar{K}_{n}+\cosh^{2} \! \big(y-\frac{\pi\tau}{\beta\ell}\big)}}\\
        &-\frac{\zeta}{\sqrt{\bar{Q}_{n}+\cosh^{2} \! \big(y-\frac{\pi\tau}{\beta\ell}\big)}}\Bigg]
    \end{split}
\label{eq:geon-F-correction}
\end{align}
where
\begin{align}
    \begin{split}
        \bar{K}_n&:=\frac{R^{2}}{R^{2}-r_{h}^2}\sinh^{2} \! \bigg[\bigg(n+\frac{1}{2}\bigg)\pi\frac{r_{h}}{\ell}\bigg],\\
        \bar{Q}_n&:=K_{n}+\frac{r_{h}^2}{R^{2}-r_{h}^{2}}.
    \end{split}
\end{align}
Note that the transition rate now depends on the switch-off moment~$\tau$. If the detector is switched off in the distant past, $\tau\rightarrow-\infty$, the geon correction term $\Delta\dot{\mathcal{F}}_{\mathbb{R}\text{P}^{2}}$ 
\eqref{eq:geon-F-correction} is small, and the response is nearly identical to that in BTZ spacetime. 
However, $\Delta\dot{\mathcal{F}}_{\mathbb{R}\text{P}^{2}}$ 
becomes significant near $\tau=0$, and it stays significant as $\tau\rightarrow\infty$. 
This is because the isometry $J$ \eqref{eq:J-action} does not commute with translations along the trajectory~\eqref{eq:static-trajectory}: among all constant $t$ surfaces in the geon's exterior, the surface $t=0$ is the only one that extends to a smooth geodesically complete surface in the geon spacetime~\cite{Louko:1998hc}.

\subsection{Swedish geon}

Since the matrices $A$ and $B$ \eqref{eq:A-B-maintext} are a pair of free generators for the fundamental group of the Swedish geon, 
the Swedish geon Wightman function takes the form \eqref{eq:wightman-imagesum}, where each $g$ is a unique string in $A$ and $B$ and their inverses, 
such that $A$ and $A^{-1}$ may not appear next to each other and $B$ and $B^{-1}$ may not appear next to each other, but subject to no other conditions. 
For $g$ with exactly $N\ge1$ generators in the string, there are $4\cdot3^{N-1}$ terms in the sum, in addition to the pure $\text{AdS}_3$ term $W_{\text{AdS}_3}(x,x')$; 
for $g$ with up to $K\ge1$ generators in the string, there are 
\begin{align}
    \sum_{N=1}^{K}4\cdot3^{N-1}=2(3^{K}-1)
\end{align}
terms in addition to the pure $\text{AdS}_3$ term. 

The detector's transition rate is now obtained 
from 
\eqref{eq:transitionrate} with
\eqref{eq:static-trajectory}, 
\eqref{eq:wightman-imagesum} and~\eqref{eq:AdS-wightman}. 

In numerical evaluation, we group the terms in the sum by the number of generators in the string, and we truncate the sum at a finite number $K$ of generators, monitoring the accuracy of the truncation by increasing~$K$. As terms with a large number of generators are suppressed by successive factors that are negative exponentials in~$\sqrt{M}$, 
accurate evaluation requires fewer terms for larger values of~$M$.

\section{Results}
\label{sec:results}

\begin{figure}[t]
    \centering
    \includegraphics[width=\columnwidth]{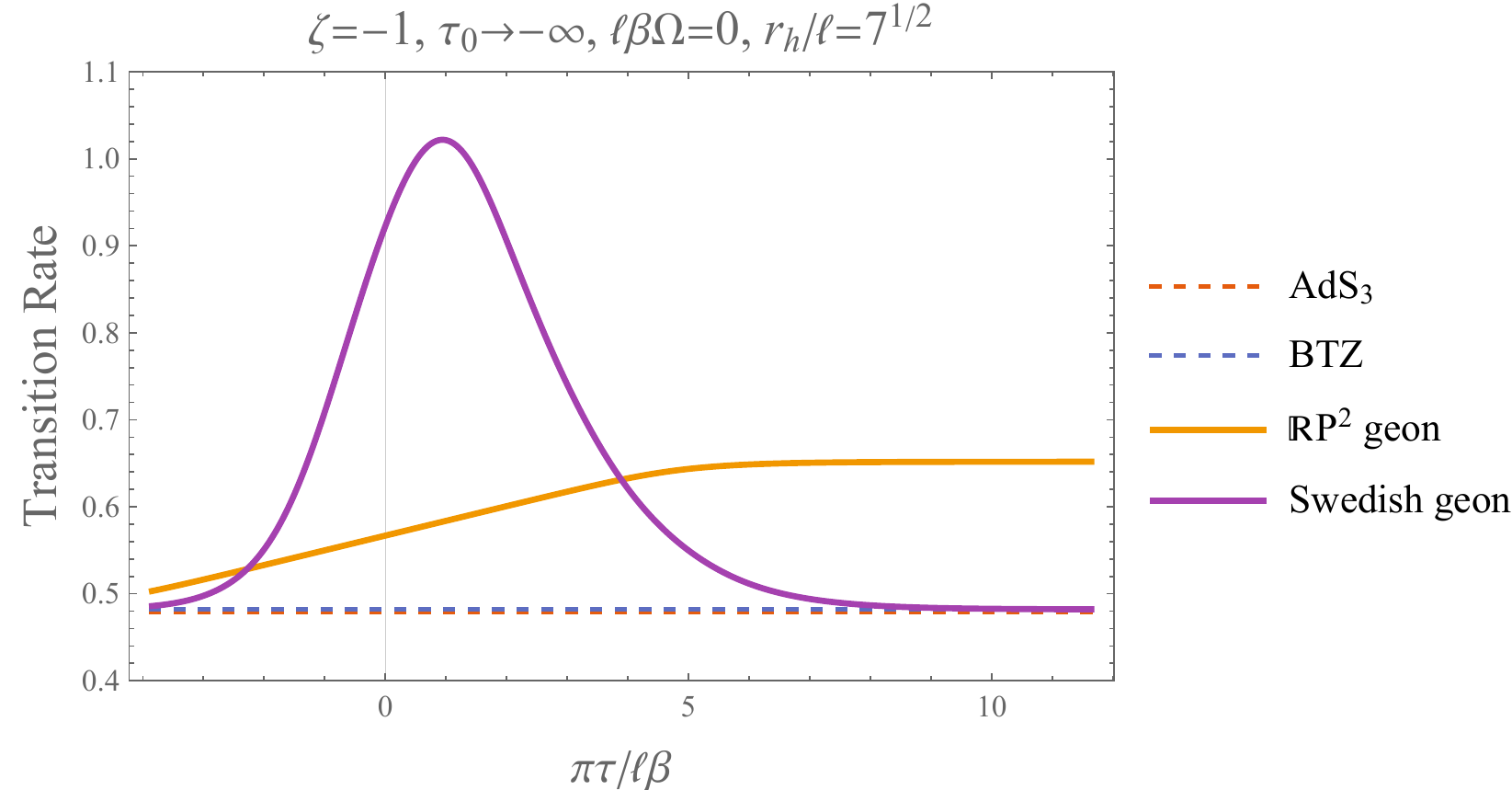}\\
    (a) \\[5ex]
    \includegraphics[width=\columnwidth]{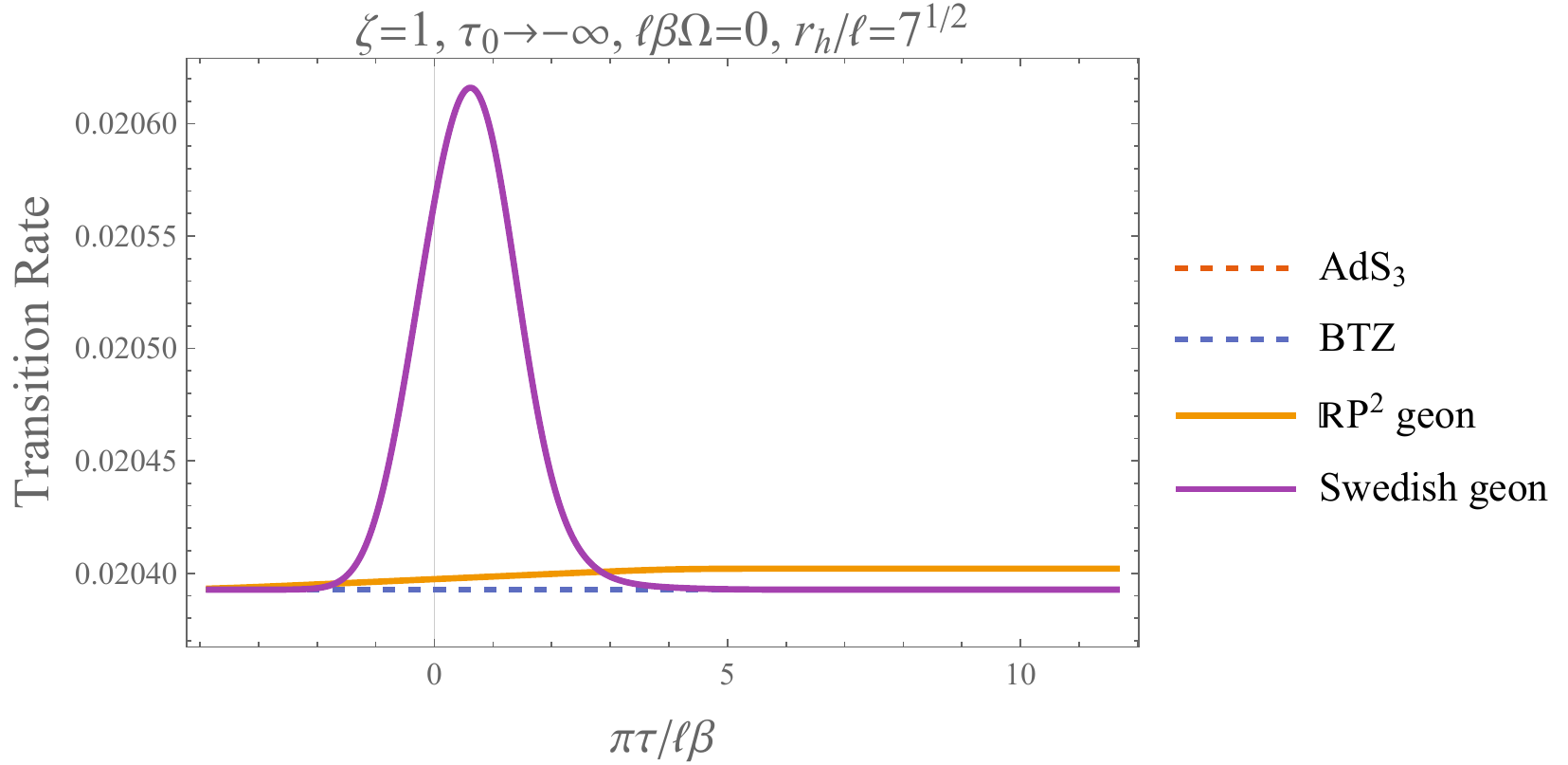}\\
    (b) \\[5ex]
    \includegraphics[width=\columnwidth]{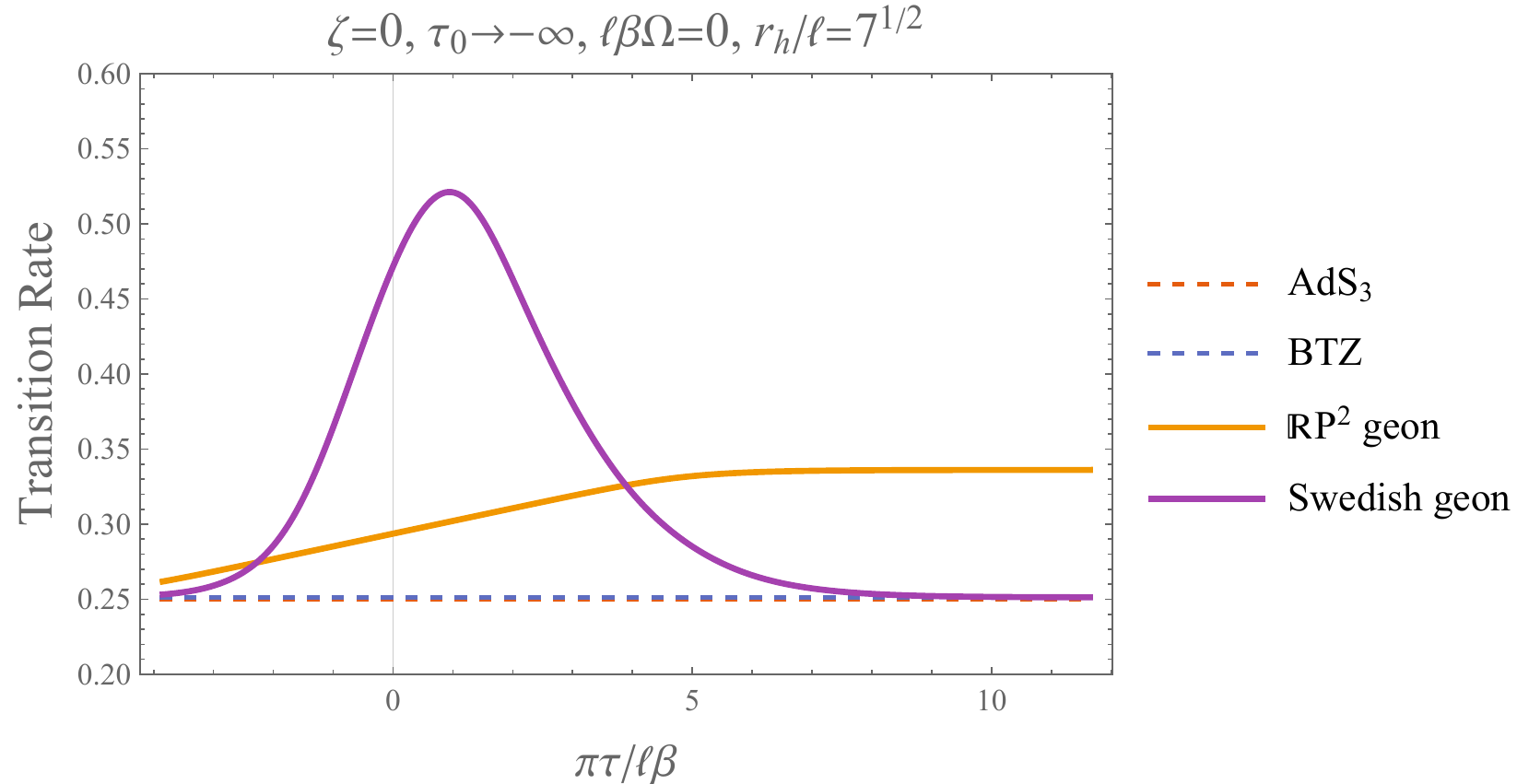}\\
    (c) 
    \caption{The transition rate of a gapless detector, switched on in the asymptotic past, as a function of the switch-off moment, for  AdS${}_3$, the BTZ black hole, the $\mathbb{R}\text{P}^{2}$ geon, and the Swedish geon. 
    The parameters are $r_{h}/\ell=\sqrt{7}$ and $R^{2}/r_{h}^{2}=\frac{25}{7}$. 
    Panels (a), (b) and (c) are for the Neumann ($\zeta=-1$),  Dirichlet ($\zeta=1$), 
    and transparent ($\zeta=0$) boundary 
    condition, respectively. 
    The transition rates on AdS${}_3$ and the BTZ black hole so close that they are indistinguishable by the naked eye.}  
    \label{fig:gapless-rates-threeconditions}
\end{figure}

Figure \ref{fig:gapless-rates-threeconditions} 
shows numerical results for the detector's transition rate as a function of the switch-off moment on AdS${}_3$, the BTZ black hole, the $\mathbb{R}\text{P}^{2}$ geon, and the Swedish geon. 
The detector is switched on in the asymptotic past and the detector's gap is set to zero. 
We have set $M=7$, 
so that $r_{h}=\sqrt{7} \ell$, 
and the detector is placed at $R=\frac{5}{\sqrt{7}} r_h$. 
The three panels show the boundary condition at infinity being respectively Neumann, Dirichlet, and transparent. 

\begin{figure}[t]
    \centering
    \includegraphics[width=\columnwidth]{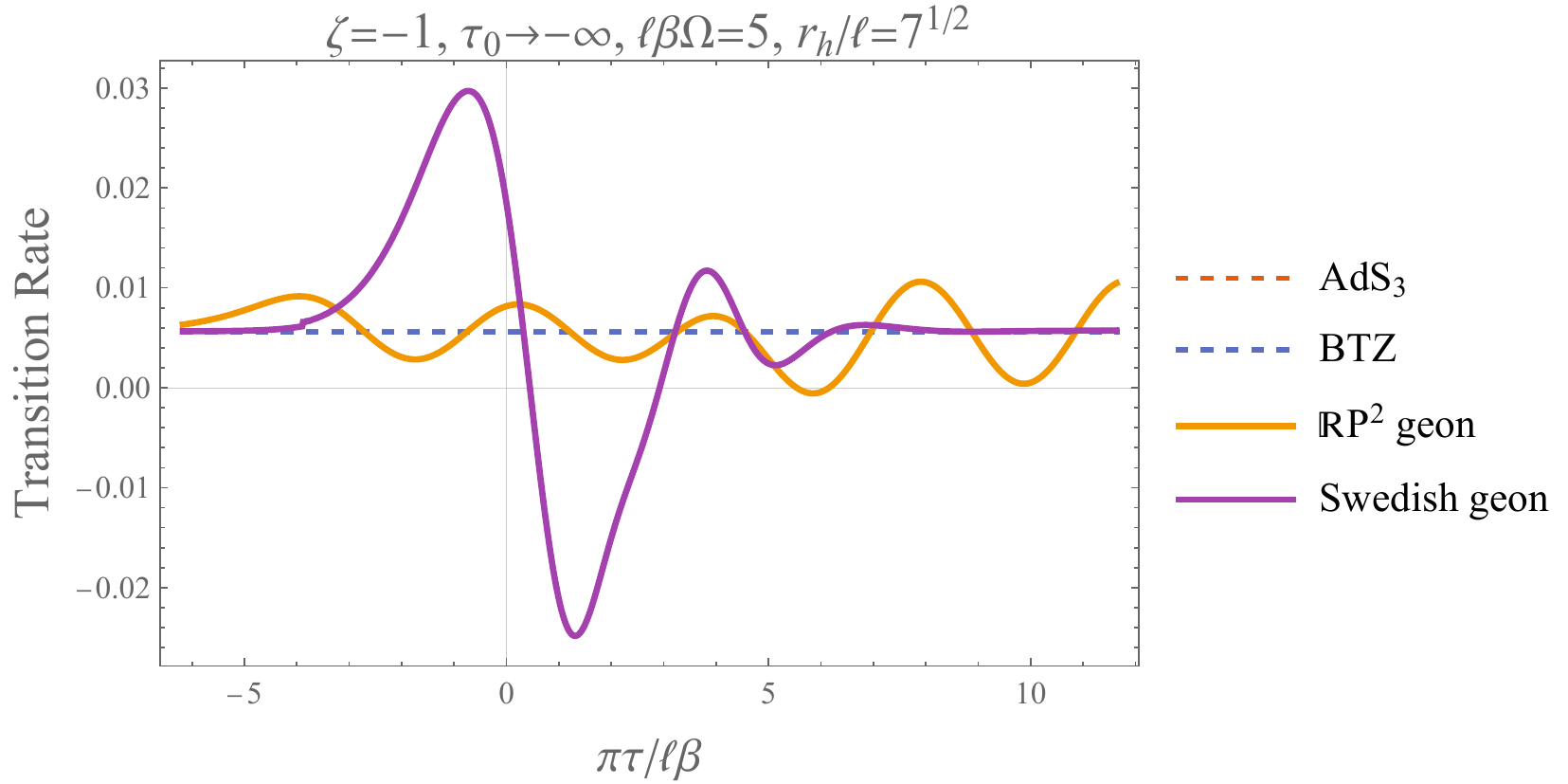}\\
    \caption{As in Figure \ref{fig:gapless-rates-threeconditions}(a) but with $\ell\beta\Omega=5$, giving the excitation rate of a detector with a nonzero gap. The transition rates on AdS${}_3$ and the BTZ black hole are again so close that they are indistinguishable by the naked eye.} 
    \label{fig:Gapped transition rates Neumann}
\end{figure}

\begin{figure}[t]
    \centering
    \includegraphics[width=\columnwidth]{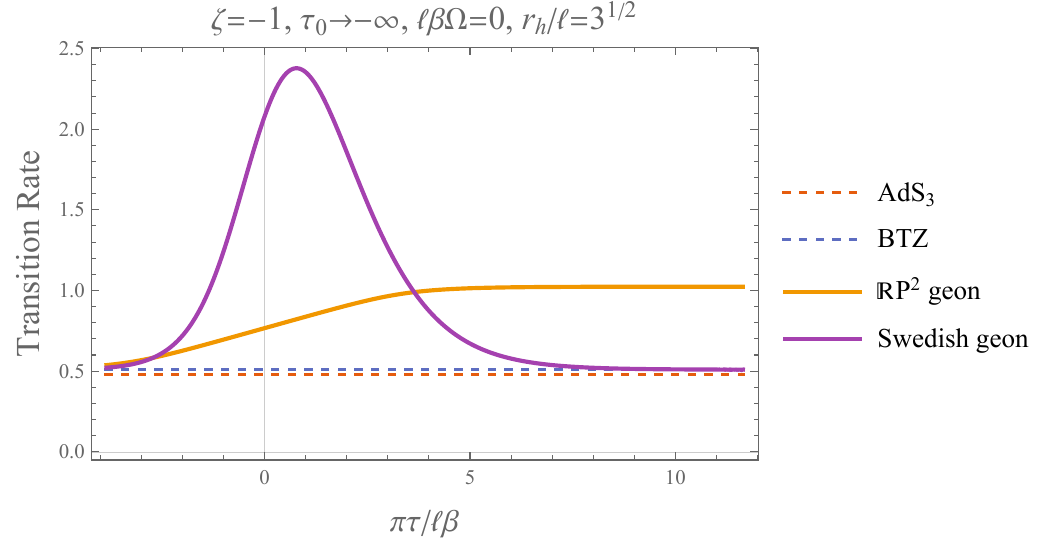}\\
    (a) \\[5ex]
    \includegraphics[width=\columnwidth]{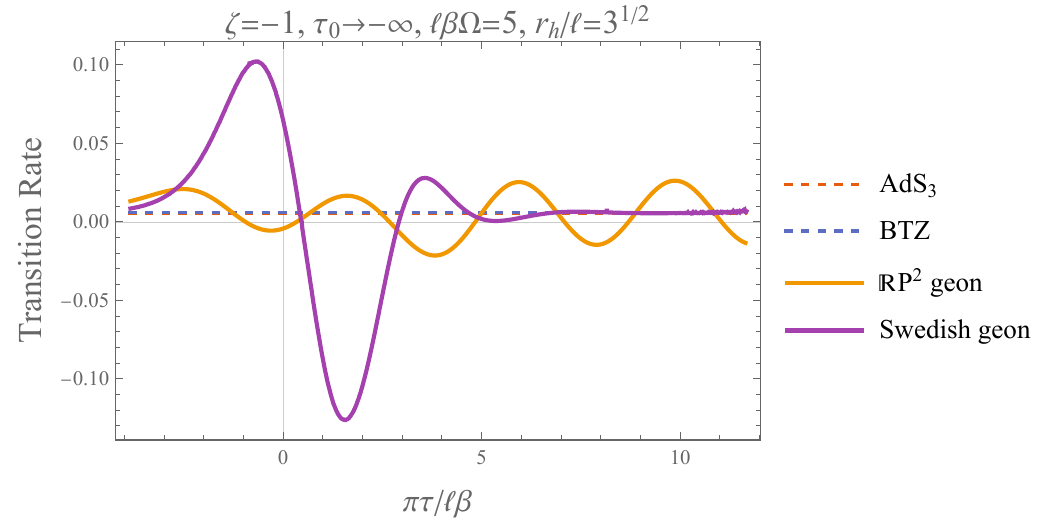}\\
    (b) 
    \caption{As in Figures \ref{fig:gapless-rates-threeconditions}(a) (Panel (a)) and \ref{fig:Gapped transition rates Neumann} (Panel (b)) 
    but with a smaller black hole, $r_{h}/\ell=\sqrt{3}$, with detector position still such that $R^2/r_{h}^{2}=\frac{25}{7}$.}  
    \label{fig:rates-smallmass}
\end{figure}

On AdS${}_3$ and on the BTZ hole, the transition rate is independent of the switch-off moment, by the time translation symmetry of the global vacuum on AdS, and by the corresponding time translation symmetry inherited by the Hartle-Hawking state on the BTZ black hole. This independence is clear in all three panels in Figure \ref{fig:gapless-rates-threeconditions}. With the parameters chosen, the transition rates on AdS${}_3$ and the BTZ hole are so close that they are indistinguishable by the naked eye. 
On the $\mathbb{R}\text{P}^{2}$ geon, the transition rate in the asymptotic past agrees with that for  the BTZ black hole, 
but in the asymptotic future it differs: 
this phenomenon was previously noticed in~\cite{Smith_2014}, and traced to the fact that the Hartle-Hawking-type state on the $\mathbb{R}\text{P}^{2}$ geon exterior is not invariant under the exterior Killing time translations, and to the way in which the late time additional contributions arise from the detector having operated in the distant past. Related observations in flat spacetimes with similar isometries were made in \cite{Langlois:2005nf,Langlois:2005if}. 

Now, for the Swedish geon, Figure \ref{fig:gapless-rates-threeconditions} shows that the transition rate agrees with that for the BTZ black hole both in the asymptotic past and in the asymptotic future, but it displays a distinct bump at intermediate times. The behaviour seen in the asymptotic past is as expected, stemming from the lack of exterior time translation symmetry of the Hartle-Hawking type state induced from the global vacuum. 
The behaviour seen in the asymptotic future is numerical evidence that the Swedish geon correlations between the asymptotic future and asymptotic past are significantly more involved than those on the $\mathbb{R}\text{P}^{2}$ geon. 
An analytic explanation of this phenomenon and the underlying correlations would require a deeper geometric study of the discrete group action by which the Swedish geon is quotiented from AdS${}_3$. 

For a detector with a nonzero gap, a plot for $\ell\beta\Omega=5$ and the Neumann boundary condition is shown in Figure~\ref{fig:Gapped transition rates Neumann}. The graph shows that the transition rate on the $\mathbb{R}\text{P}^{2}$ geon and the Swedish geon display more oscillations as a function of the switch-off moment. For the $\mathbb{R}\text{P}^{2}$ geon these oscillations continue to asymptotically late times, as noted in~\cite{Smith_2014}, but on the Swedish geon we see the late time transition rate again  asymptotically approaches  the stationary BTZ value. 

Figure \ref{fig:rates-smallmass} is as in Figures \ref{fig:gapless-rates-threeconditions}(a) and \ref{fig:Gapped transition rates Neumann} but with a smaller black hole. The qualitative features are similar, but the bumps in the Swedish geon transition rate are now noticeably larger. The AdS${}_3$ transition rate and the BTZ black hole transition rate are now distinguishable by   eye in Figure~\ref{fig:rates-smallmass}. 

Regarding numerical accuracy, the more massive Swedish geon plots include contributions from image terms with up to $K=5$ generators in the strings that define the group elements. We checked that the results did not change visibly on taking $K=6$. We also made additional spot checks at late times with $K=8$, involving a total of $13,120$ image terms. While we found this to be sufficient for the more massive black hole mass $M=7$ used in the plots, numerical accuracy for smaller values of $M$ would require more image terms. The less massive ($M=3$) Swedish geon plots include contributions from image terms with up to $K=8$ generators, for sufficient accuracy.

\section{Conclusions and discussion}
\label{sec:conclusions}

We have considered the transition rate of a static UDW detector in two $(2+1)$-dimensional 
black hole spacetimes that are 
isometric to the static BTZ black hole outside the horizon but have the spatial topology of a topological geon, with no second exterior region behind the black-and-white-hole horizons. The detector was coupled linearly to a conformal scalar field, prepared in the Hartle-Hawking-type state that is induced from the global vacuum on AdS${}_3$, with Dirichlet, Neumann or transparent boundary conditions at the infinity. The detector was treated in first-order perturbation theory and assumed switched on in the asymptotic past. 

The spacetimes were the $\mathbb{R}\text{P}^{2}$ geon \cite{Louko:1998hc} and the Swedish geon of \AA{}minneborg \emph{et al\/}~\cite{aaminneborg_1998}, each obtained as a quotient of an open subset in AdS${}_3$ under a discrete isometry group. 
For both of these topological geon spacetimes, we found that the detector's response differs from that for the BTZ black hole, and the responses for the two spacetimes differ from each other. 
In particular, the detector's late time transition rate in the $\mathbb{R}\text{P}^{2}$ geon differs from that in the BTZ black hole, as previously found in~\cite{Smith_2014}, and the geometric reasons for this were identified in a related context in \cite{Langlois:2005nf,Langlois:2005if}; by contrast, our numerical results show that the late time response in the Swedish geon approaches that on the BTZ black hole. An analytic explanation of this late time phenomenon would require a deeper geometric study of the discrete group action by which the Swedish geon is quotiented from AdS${}_3$. 

Our numerical results also show that, for the Swedish geon, at intermediate times the transition rate is much larger than for the $\mathbb{R}\text{P}^{2}$ geon with the same mass. This is perhaps not unexpected, as the image sum for the field's Wightman function on the $\mathbb{R}\text{P}^{2}$ geon consists of a strict subset of the terms in the image sum on the Swedish geon, but a more quantitative explanation would again require a deeper study of the discrete group action by which the Swedish geon is constructed from AdS${}_3$. 

The salient lesson from our results is that specific information about the interior topology of a black hole is  accessible to an  observer outside the black hole  by allowing them  to probe a quantum state that has been prepared with knowledge of the global topology of the spacetime. 

A limitation of our results is the restriction to static observers. The response rate of infalling detectors for the spinless \cite{Preciado-Rivas:2024gzm} and spinning \cite{Wang:2024zny} BTZ black holes has been recently investigated, as well as for  the $\mathbb{R}\text{P}^{2}$ geon ~\cite{Spadafora:2024gqj}. These results suggest that
the Swedish geon case will exhibit 
more detailed differences as compared to its 
$\mathbb{R}\text{P}^{2}$ geon counterpart.
This remains an interesting subject for future work.

\acknowledgments

We thank Kirill Krasnov and Fredrik Str\"omberg for discussions on fundamental domains
and an anonymous referee for helpful presentational suggestions. 
This work was supported in part by the Natural Sciences and Engineering Research Council of Canada.
The work of JL was supported by United Kingdom Research and Innovation Science and Technology Facilities Council [grant numbers ST/S002227/1, ST/T006900/1 and ST/Y004523/1].
For the purpose of open access, the authors have applied a CC BY public copyright licence to any Author Accepted Manuscript version arising.

\begin{widetext}

% \widetext 

\appendix 

\section{Swedish geon generators\label{sec:generators}}

In this appendix we construct a convenient pair of free generators of the isometry group by which the Swedish geon is quotiented from AdS${}_3$. 
We first consider the spacelike hypersurface on which the geometry is locally that of the Poincar\'e disk, and we then extend the generators to the spacetime. 

\subsection{Fundamental domain on the Poincar\'e disk}

Recall that the Poincar\'e metric on the open unit disk $|z|<1$ of $\mathbb{C}$ is 
\begin{align}
ds^2 = \frac{4 |dz|^2}{{(1 - |z|^2)}^2} 
\, . 
\label{eq:diskmetric}
\end{align}
The metric has constant negative curvature, with Ricci scalar~$-2$. 
The metric is invariant under the fractional linear $SU(1,1)$ action 
\begin{align}
z \mapsto U(z)=\frac{\alpha z+\overline{\beta}}{\beta z+\overline{\alpha}}, 
\end{align}
where $U \in SU(1,1)$ is written as 
\begin{align}
        U=\begin{pmatrix}
            \alpha&\overline{\beta}\\
            \beta&\overline{\alpha}
        \end{pmatrix},
\label{eq:SU11-parametrisation}
\end{align}
where $\alpha$ and $\beta$ are complex numbers with 
$|\alpha|^{2}-|\beta|^{2}=1$. 
The only degeneracy in this action is that $U$ and $-U$ give the same fractional linear transformation. The isometry group of the Poincar\'e disc is hence $PSU(1,1)$, and elements in $PSU(1,1)$ may be represented as in \eqref{eq:SU11-parametrisation} when understood modulo overall sign. 

To establish notation, we define three one-parameter families of $SU(1,1)$ matrices by 
\begin{subequations}
\label{eq:R-DH-DV-defs}
\begin{align}
        R(\theta)&:=\begin{pmatrix}
            e^{i\theta/2}&0\\
            0&e^{-i\theta/2}
        \end{pmatrix},
\\[1ex]
D_{H}(\psi)&:=\begin{pmatrix}
            \cosh(\psi/2)&\sinh(\psi/2)\\
            \sinh(\psi/2)&\cosh(\psi/2)
        \end{pmatrix},
\label{eq:DH-def}
\\[1ex]
D_{V}(\psi)&:=\begin{pmatrix}
            \cosh(\psi/2)&i\sinh(\psi/2)\\
            -i\sinh(\psi/2)&\cosh(\psi/2)
        \end{pmatrix}. 
\end{align}
\end{subequations}
The action of $R(\theta)$ on the disk is a rotation about the origin, and $\theta$ is the rotation angle, measured positive counterclockwise. 
The action of $D_{H}(\psi)$ is a boost with fixed points at $z=\pm1$, such that positive values of the rapidity $\psi$ move points to the right, towards $z=1$. 
The action of $D_{V}(\psi)$ is a boost with fixed points at $z=\pm i$, such that positive values of the rapidity $\psi$ move points upwards, towards $z=i$. The subscripts $H$ and $V$ stand respectively for ``horizontal'' and ``vertical.'' 

With this notation, let $a>2\arsinh1$, and let 
\begin{subequations}
\label{eq:poincareAB-symm}
\begin{align}
{\tilde A} &:= D_{H}(a),
\\[1ex]
{\tilde B} &:= D_{V}(a). 
\end{align}
\end{subequations}
${\tilde A}$ and ${\tilde B}$ act on the disk as illustrated in Figure~\ref{fig:poincarediskdomains}(a), mapping the four symmetrically-chosen geodesic circle segments to each other as shown. The condition $a>2\arsinh1$ is the condition for the circle segments not to intersect, neither on the open disk nor on its boundary. From the figure it is hence clear that the group generated by ${\tilde A}$ and ${\tilde B}$ acts on the disk freely and properly discontinuously, and the domain between the four circle segments in the figure serves as a fundamental domain for the group action. This is the construction given in \cite{aaminneborg_1998}.

\begin{figure}[t]
\centering
\raisebox{3.5ex}[0pt][0pt]{\includegraphics[width=8 cm]{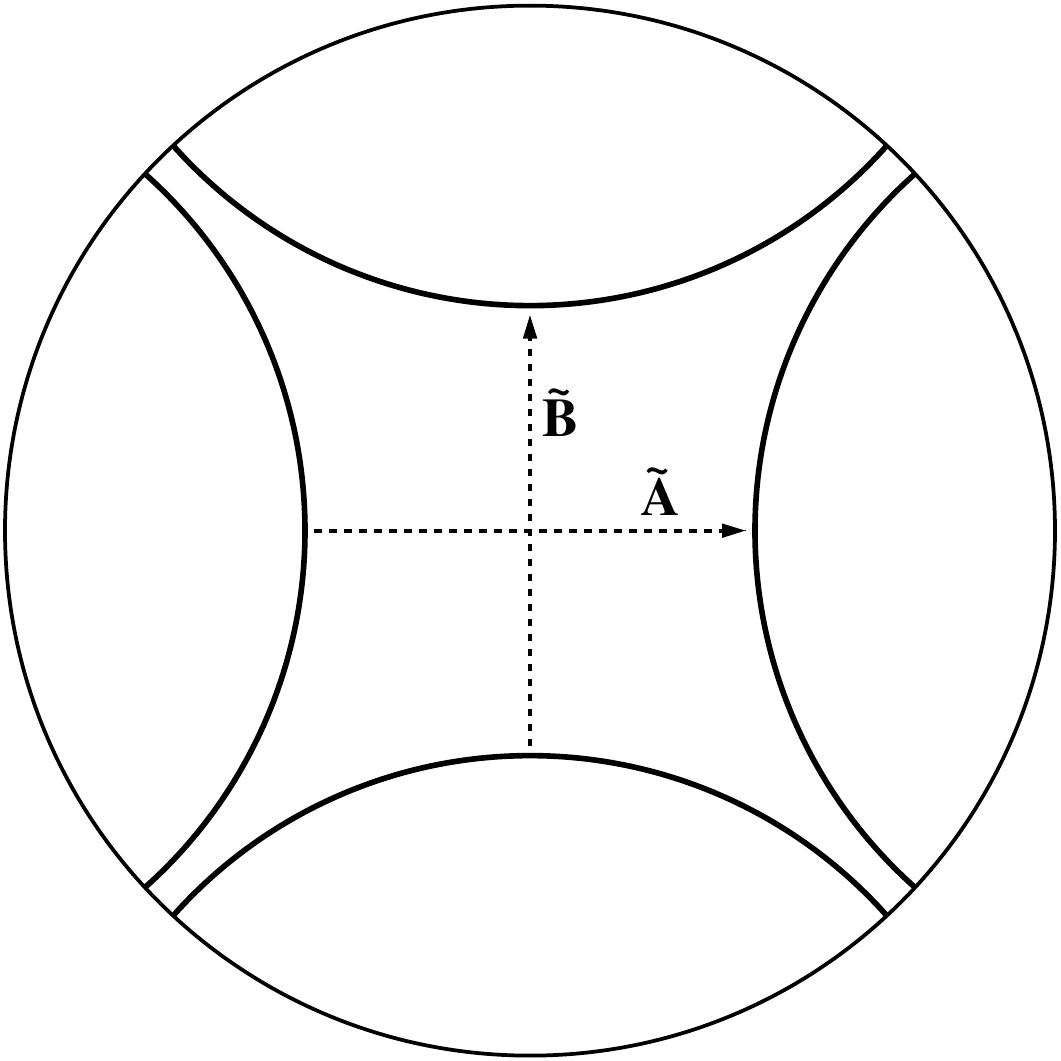}}
\hspace{1ex}
\includegraphics[width=9.2 cm]{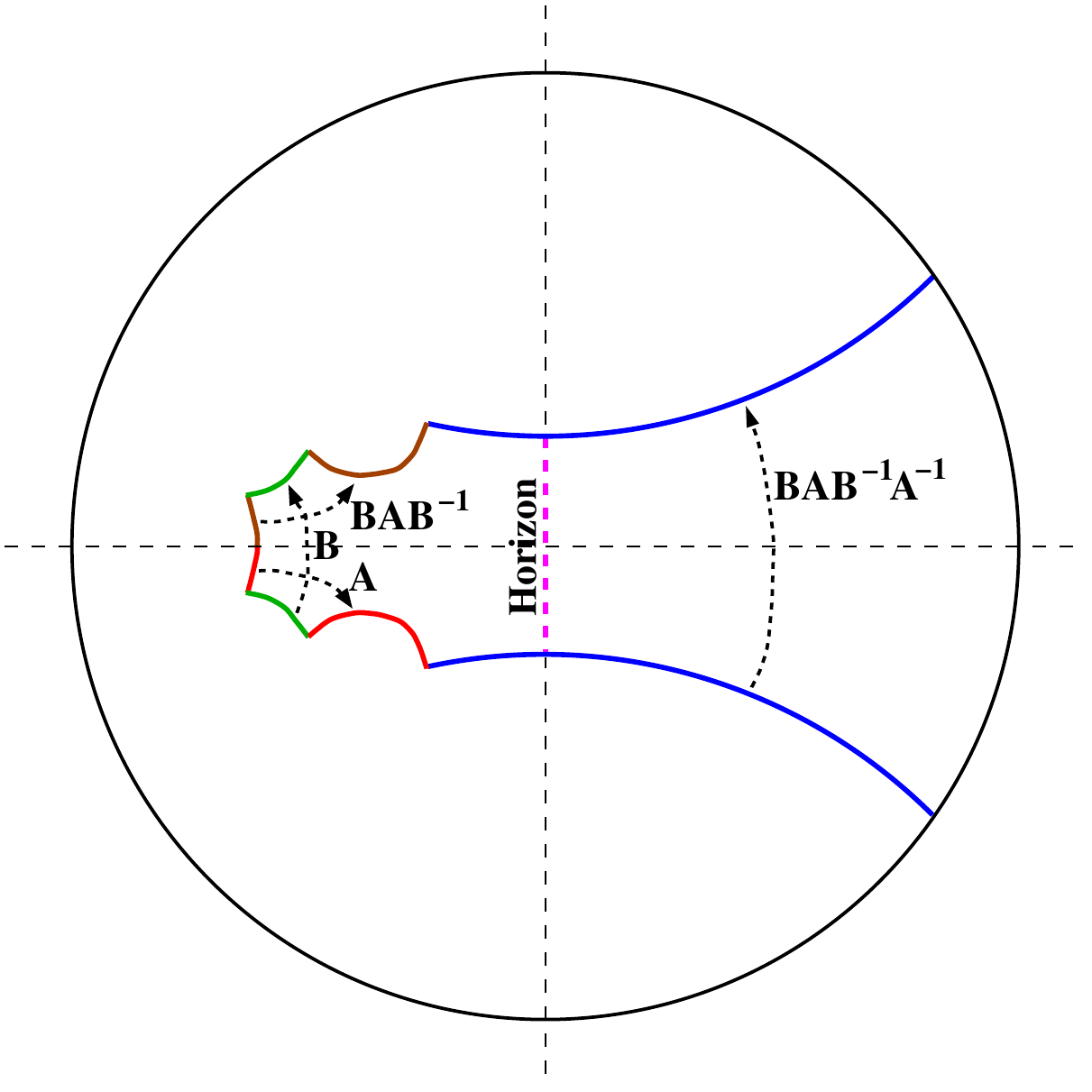}

(a) \hspace{7.9cm} (b) 
\caption{Two fundamental domains on the Poincar\'e disk. {\bf (a)} In panel~(a), the fundamental domain is between the four symmetrically-arranged geodesic circle segments, and the identifications of the segments use the generators ${\tilde A}$ and ${\tilde B}$ \eqref{eq:poincareAB-symm} as shown. 
The fundamental domain reaches the boundary of the disk in four segments, and the identifications join these segments into a circle. 
{\bf (b)} In panel~(b), the fundamental domain is between the seven geodesic circle segments, and the identifications of the segments use the generators $A$ and $B$ \eqref{eq:poincareAB} as shown. The fundamental domain is adapted to the infinity, where it reaches the boundary of the disk in a single segment, and the identification by 
$B A B^{-1} A^{-1} = - D_V(\psi_h)$, with $\psi_h$ given by 
\eqref{eq:psi-h-def}, joins this segment into a circle. 
The horizon is on the axis of $D_V(\psi_h)$ as shown.}
\label{fig:poincarediskdomains}
\end{figure}

The symmetric fundamental domain shown in Figure \ref{fig:poincarediskdomains}(a) intersects the boundary of the disk in four segments, but it is clear from the identifications that these four segments connect to form a circle, and the quotient space hence has a single trumpet-like infinity. 
A fundamental domain adapted to the infinity is displayed in Figure~\ref{fig:poincarediskdomains}(b), 
where the identifications of the boundary segments are as shown, 
and the $SU(1,1)$ matrices $A$ and $B$ are given by 
\begin{subequations}
\label{eq:poincareAB}
\begin{align}
A &= R(q) D_H(-\chi_A) D_V(-a) D_H(\chi_A) R(-q)
\,,
\\ 
B &= D_H(-\chi_B) D_V(a)  D_H(\chi_B)
\,,
\end{align}
\end{subequations}
with 
\begin{subequations}
\begin{align}
q &=  \arctan \! \left(\frac{\sqrt{S^4-1}}{S}\right)
\,,
\label{eq:q-def}
\\
\chi_A &= \artanh \! \left( \frac{S^2}{\sqrt{S^4 + S^2 - 1}} \right) 
\,,
\\ 
\chi_B &= \artanh \! \left(\frac{1}{S}\right)
\,, 
\end{align}
\end{subequations}
where 
\begin{align}
S := \sinh(a/2), \ \ \  C := \cosh(a/2) \,.
\label{eq:S-C-defs}
\end{align}
The map between the two infinity-reaching boundary segments is by 
\begin{align}
B A B^{-1} A ^{-1} = - D_V(\psi_h)
\,,
\label{eq:horgen-poincare}
\end{align}
where 
\begin{align}
\psi_h = 4\arcosh \! \left(S^2\right)
\,. 
\label{eq:psi-h-def}
\end{align}

The transformations leading from Figure \ref{fig:poincarediskdomains}(a) with \eqref{eq:poincareAB-symm}
to Figure \ref{fig:poincarediskdomains}(b) with \eqref{eq:poincareAB}
are given in Appendix~\ref{sec:deformation}. 

In Appendix \ref{sec:topology} it is shown that the quotient surface is a punctured torus, and $A/\{\pm1\}$ and $B/\{\pm1\}$ are a pair of free generators of the fundamental group of this surface.

\subsection{Generators in the spacetime}

Recall that we realise AdS${}_3$ as the hyperboloid \eqref{eq:embedding-hyperboloid} 
in $\mathbb{R}^{2,2}$ with the metric~\eqref{eq:2+2embeddingmetric}. 
We write the action of the $O(2,2)$ isometry group 
on AdS${}_3$ 
in terms of the embedding space coordinates as 
\begin{align}
\begin{pmatrix}
T_2\\
T_1\\
X_2\\
X_1
\end{pmatrix}
\mapsto 
\undertilde U 
\begin{pmatrix}
T_2\\
T_1\\
X_2\\
X_1
\end{pmatrix} \,, 
\label{eq:O22rep}
\end{align}
where $\undertilde U$ denotes $O(2,2)$ elements in the defining matrix representation. 

In terms of this realisation, we embed the Poincar\'e disc of Figure \ref{fig:poincarediskdomains}(b) in AdS${}_3$ by 
\begin{subequations}
\begin{align}
T_2 &= 0 \,,
\\
\frac{T_1}{\ell} &= \frac{1 + {|z|}^2}{1 - {|z|}^2}\,, 
\\
\frac{X_2 + i X_1}{\ell} &= \frac{2z}{1 - {|z|}^2}\,, 
\end{align}
\end{subequations}
as the spacelike hypersurface $T_2=0$ with $T_1>0$. 
The induced metric on this hypersurface is $\ell^2$ times~\eqref{eq:diskmetric}. 
The isometries of the Poincar\'e disk by the $SU(1,1)$ matrices \eqref{eq:R-DH-DV-defs} are induced from the AdS${}_3$ isometries by the respective $O(2,2)$ matrices 
\begin{subequations}
\begin{align}
        \undertilde R(\theta)&=\begin{pmatrix}
            1&0&0&0\\
            0&1&0&0\\
            0&0&\cos\theta&-\sin\theta\\
            0&0&\sin\theta&\cos\theta
        \end{pmatrix},\\
        \undertilde {D_H}(\psi)&=\begin{pmatrix}
            1&0&0&0\\
            0&\cosh\psi&\sinh\psi&0\\
            0&\sinh\psi&\cosh\psi&0\\
            0&0&0&1
        \end{pmatrix},\\
        \undertilde {D_V}(\psi)&=\begin{pmatrix}
            1&0&0&0\\
            0&\cosh\psi&0&\sinh\psi\\
            0&0&1&0\\
            0&\sinh\psi&0&\cosh\psi
        \end{pmatrix}.
\end{align}
\end{subequations}
It can then be verified that the Poincar\'e disc isometries by $A$ and $B$ \eqref{eq:poincareAB} are induced from the respective AdS${}_3$ isometries by 
\begin{subequations}
\label{eq:Atilde-Btilde}
\begin{align}
        \undertilde A&=
        \begin{pmatrix}
            1&0&0&0\\[1ex]
            0&{\displaystyle\frac{\cosh(6z)+2\cosh(4z)+\cosh(2z)}{4 \sinh^2(z)}}&{\displaystyle\frac{\sqrt{\cosh(2z)}\cosh(4z)}{\sinh^2(z)}}&\sinh(4z)\\[3ex]
            0&{\displaystyle-\frac{\sqrt{\cosh(2z)}}{\sinh^2(z)}}&- \coth^2(z)&0\\[3ex]
            0&{\displaystyle-\frac{\cosh(5z)+3\cosh(3z)+4\cosh(z)}{2\sinh(z)}}&-4[\cosh(2z)]^{3/2}\coth(z)&-\cosh(4z)
        \end{pmatrix},\\[2ex]
        \undertilde B&=
        \begin{pmatrix}
            1&0&0&0\\[1ex]
            0&{\displaystyle\frac{\cosh(4z)+\cosh(2z)}{2 \sinh^2(z)}}&{\displaystyle\frac{[\cosh(2z)]^{3/2}}{\sinh^2(z)}}&{\displaystyle\frac{\cosh(3z)+\cosh(z)}{\sinh(z)}}\\[3ex]
            0&{\displaystyle-\frac{[\cosh(2z)]^{3/2}}{\sinh^2(z)}}&- \coth^2(z)&-2\sqrt{\cosh(2z)}\coth(z)\\[3ex]
            0&{\displaystyle\frac{\cosh(3z)+\cosh(z)}{\sinh(z)}}&2\sqrt{\cosh(2z)}\coth(z)
            &2\cosh(2z)+1
        \end{pmatrix},
\end{align}
\end{subequations}
where $z=\frac18 \psi_h$. 
The Poincar\'e disc isometry by $B A B^{-1} A^{-1}$ \eqref{eq:horgen-poincare} is induced from the AdS${}_3$ isometry by $\undertilde {D_V}(\psi_h)$. 
It follows that the quotient of the Poincar\'e disc evolves into a single-exterior spinless BTZ black whole with $2\pi\sqrt{M} = \psi_h$, where $M$ is the mass, and the exterior is covered by the BTZ coordinates $(t,r,\phi)$ in \eqref{eq:BTZcoords-embedding} and \eqref{eq:BTZmetric}, 
with the identification $(t,r,\phi) \sim (t,r,\phi+2\pi)$. 
In terms of~$M$, \eqref{eq:Atilde-Btilde} then has $z=\frac14 \pi\sqrt{M}$.

\section{Fundamental domain deformation\label{sec:deformation}}

In this appendix we describe the deformation between the two fundamental domains shown in Figure \ref{fig:poincarediskdomains}. 

We start from the Poincar\'e disc of Figure \ref{fig:poincarediskdomains}(a), with ${\tilde A}$ and ${\tilde B}$ given by \eqref{eq:poincareAB-symm}. 
We map the disc to the upper half-plane by the fractional linear transformation 
\begin{align}
w = V \times 
\frac{\bigl(1- i (C+S)\bigr)z + i - (C+S)}{\bigl(1+ i (C+S)\bigr)z + i + (C+S)}
\,, 
\label{eq:z-to-w-fraclin}
\end{align}
where $S$ and $C$ are given by \eqref{eq:S-C-defs}, 
and $V := \sqrt{\frac{S+1}{S-1}}$. 
Writing $w = u + iv$, where $u\in\mathbb{R}$ and $v>0$, the metric \eqref{eq:diskmetric} is mapped to the upper half-plane hyperbolic metric 
\begin{align}
ds^2 = \frac{du^2 + dv^2}{v^2} . 
\end{align}
The fractional linear transformations by ${\tilde A}$ and ${\tilde B}$ are mapped respectively to fractional linear transformations by the $SL(2,\mathbb{R})$ matrices 
\begin{subequations}
\begin{align}
    {\hat A}&=\begin{pmatrix}
            -C(S-1)&-S^2 V\\
            S^2 V^{-1}&C(S+1)
        \end{pmatrix},
\\[1ex]
    {\hat B}&=\begin{pmatrix}
            C&SV\\
            SV^{-1}&C
        \end{pmatrix}.
\end{align}
\end{subequations}
The fixed points $z=\pm1$ of ${\tilde A}$ are mapped to 
the fixed points of ${\hat A}$, which are respectively at 
$w = -V(C-1)S^{-1}$ and $w = -V(C+1)S^{-1}$, 
and the fixed points $z=\pm i$ of ${\tilde B}$ 
are mapped to the fixed points of ${\hat B}$, 
which are respectively at $w = \pm V$. 
The fundamental domain in Figure \ref{fig:poincarediskdomains}(a) is mapped to the fundamental domain $D_1$ in Figure~\ref{fig:halfplaneD1}, with the identifications described in the caption. 
Figures \ref{fig:halfplaneD1}--\ref{fig:halfplaneD2-toD3} then show how $D_1$ can be replaced by the new fundamental domain $D_3$ in Figure~\ref{fig:halfplaneD2-toD3}(b), as described in the captions. 
In particular, it follows that there is a horizon, generated by the boost 
\begin{align}
{\hat B}{\hat A}{\hat B}^{-1}{\hat A}^{-1} 
= 
- \begin{pmatrix}
            2S^4 - 1&2CS^2\sqrt{S^2-1}\\
            2CS^2\sqrt{S^2-1}&2S^4 - 1
\end{pmatrix}, 
\end{align}
which has fixed points $w = \pm1$ and the boost parameter $\psi_h$~\eqref{eq:psi-h-def}. 

Finally, we map $D_3$ in the upper half-plane back to the Poincar\'e disk by the fractional linear transformation 
\begin{align}
    z =\frac{iw+1}{-iw+1} ,
\end{align}
which maps $w=0$ to $z=1$, $w=i$ to $z=0$ and $w=\infty$ to $z=-1$. 
This gives the fundamental domain in Figure \ref{fig:poincarediskdomains}(b). 
$A$ and $B$ are boosts with rapidity~$a$, with fixed points respectively at   
\begin{subequations}
\begin{align}
&A: \ \  e^{iq}\frac{- S^2 \mp i \sqrt{S^2 - 1}}{\sqrt{S^4 + S^2 - 1}}
\,,
\\
&B: \ \ \frac{-1 \pm i\sqrt{S^2-1}}{S}\,,
\end{align}
\end{subequations}
where $q$ is given by~\eqref{eq:q-def}. From this it follows that $A$ and $B$
have the expresions given in 
\eqref{eq:poincareAB}--\eqref{eq:S-C-defs}, using the fact that $D_H(\psi)$ 
% \eqref{eq:DH-def} 
maps the fixed points $\pm i$ of $D_V(\psi)$ to respectively $\tanh\psi \pm i/\cosh\psi$.

\begin{figure}[p]
\centering
\includegraphics[width=8.0 cm]{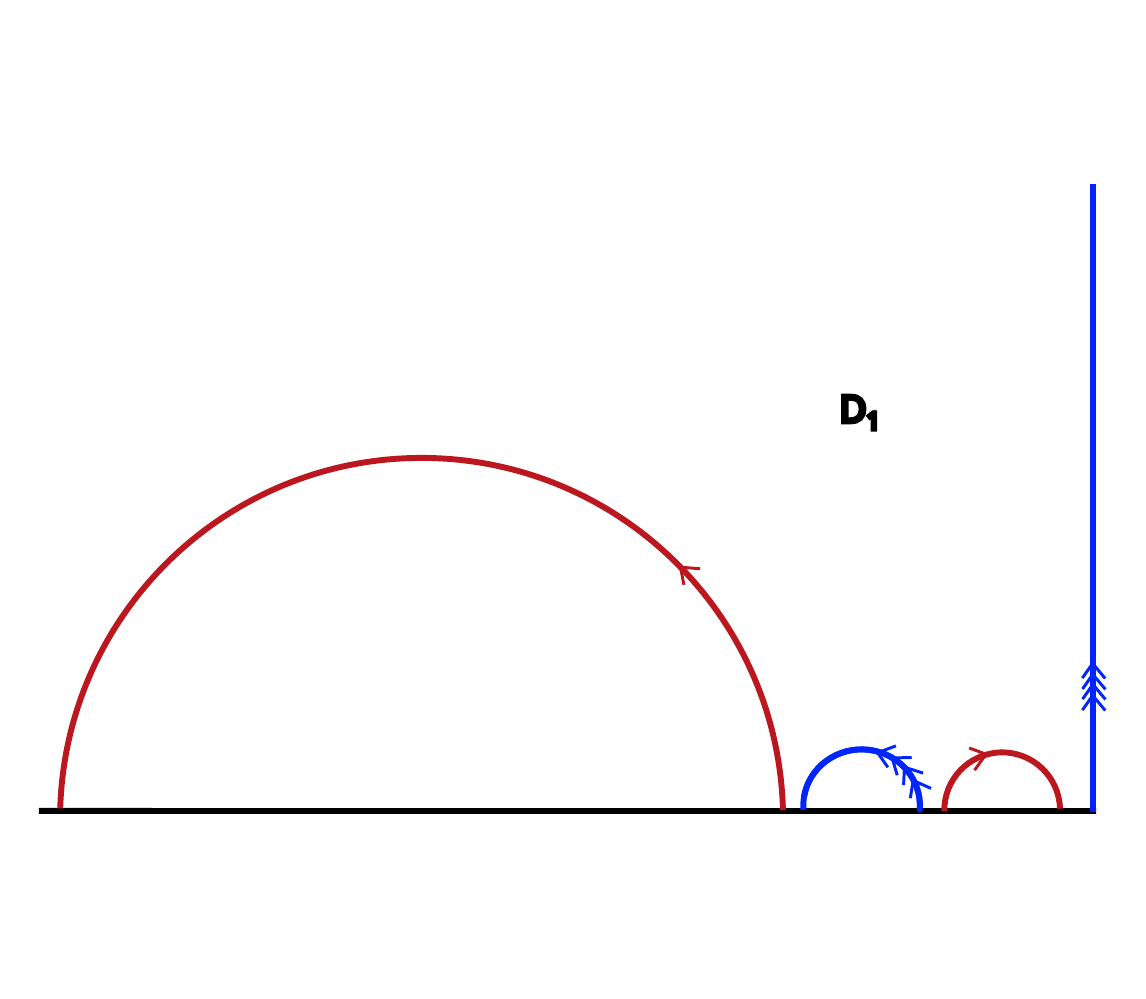}
\includegraphics[width=8.0 cm]{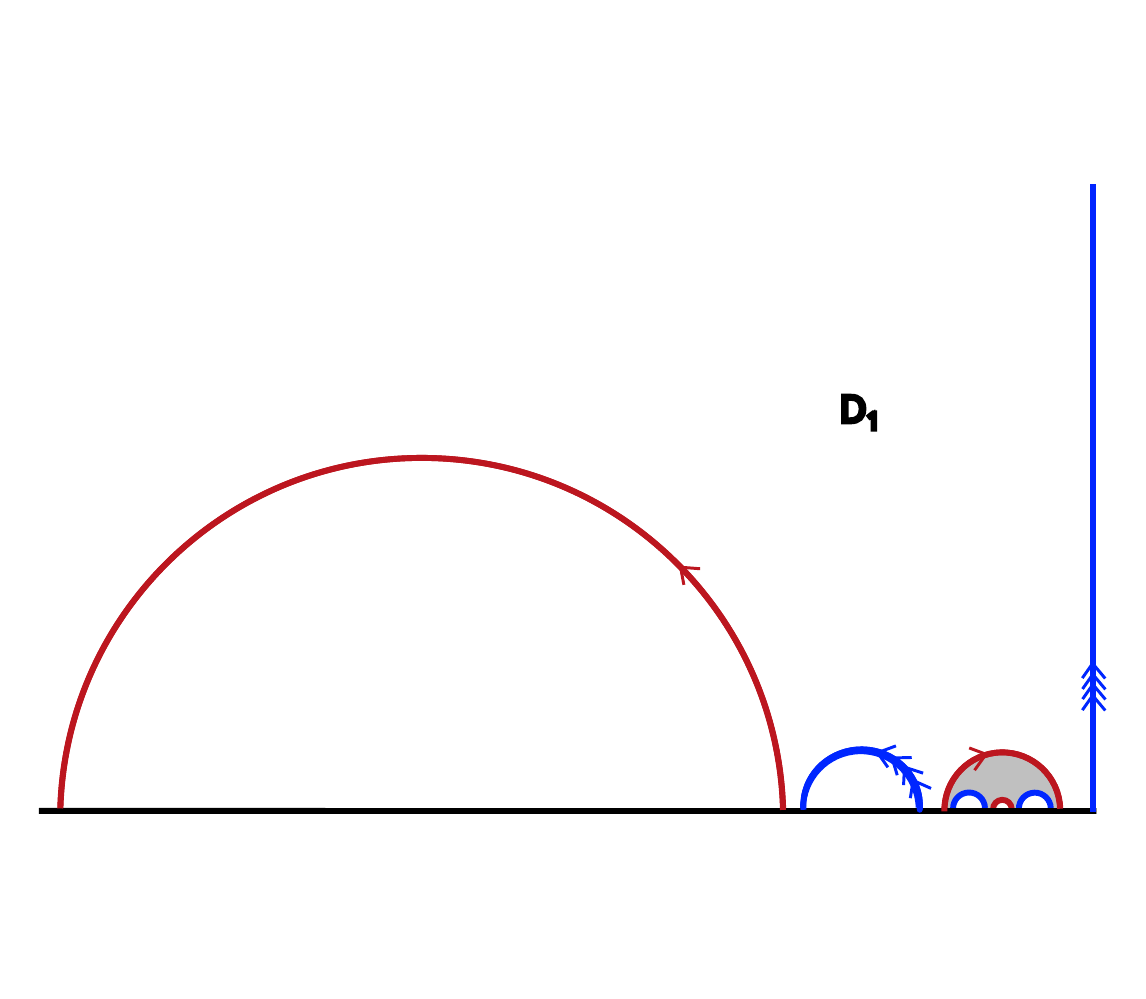}

(a) \hspace{7.5cm} (b) 

\caption{{\bf (a)} The upper half-plane fundamental domain $D_1$ to which the fundamental domain in Figure \ref{fig:poincarediskdomains}(a) is mapped by the fractional linear transformation~\eqref{eq:z-to-w-fraclin}. ${\hat A}$ maps the left single-arrowed semicircle to the right one, and ${\hat B}$ maps the quadruple-arrowed semicircle to the positive imaginary axis. 
{\bf (b)}
$D_1$ together with its image under ${\hat A}$, shown shaded.}
\label{fig:halfplaneD1}
\end{figure}

\begin{figure}[p]
\centering
\includegraphics[width=9.0 cm]{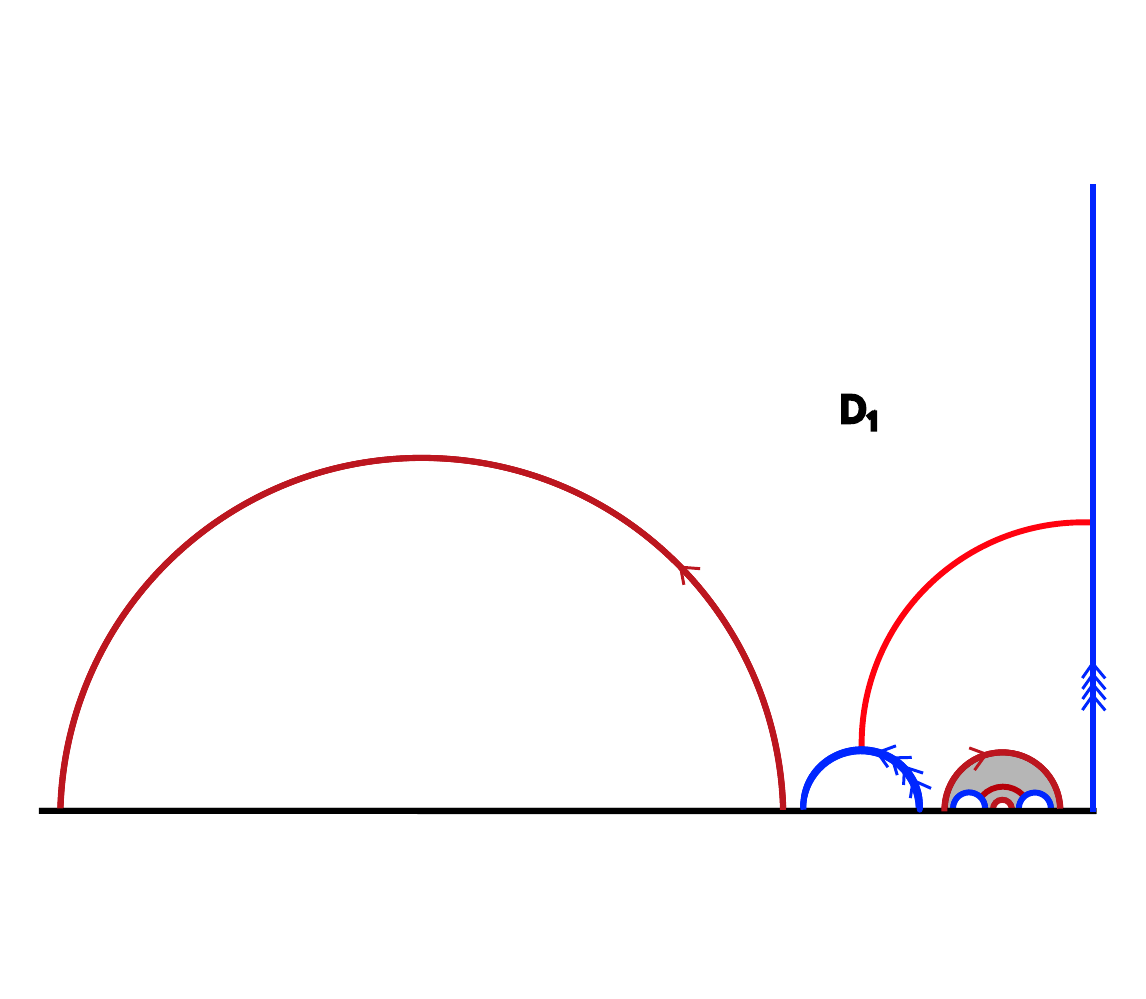} \hspace{1ex}
\raisebox{9ex}[0pt][0pt]{\includegraphics[width=7.5 cm]{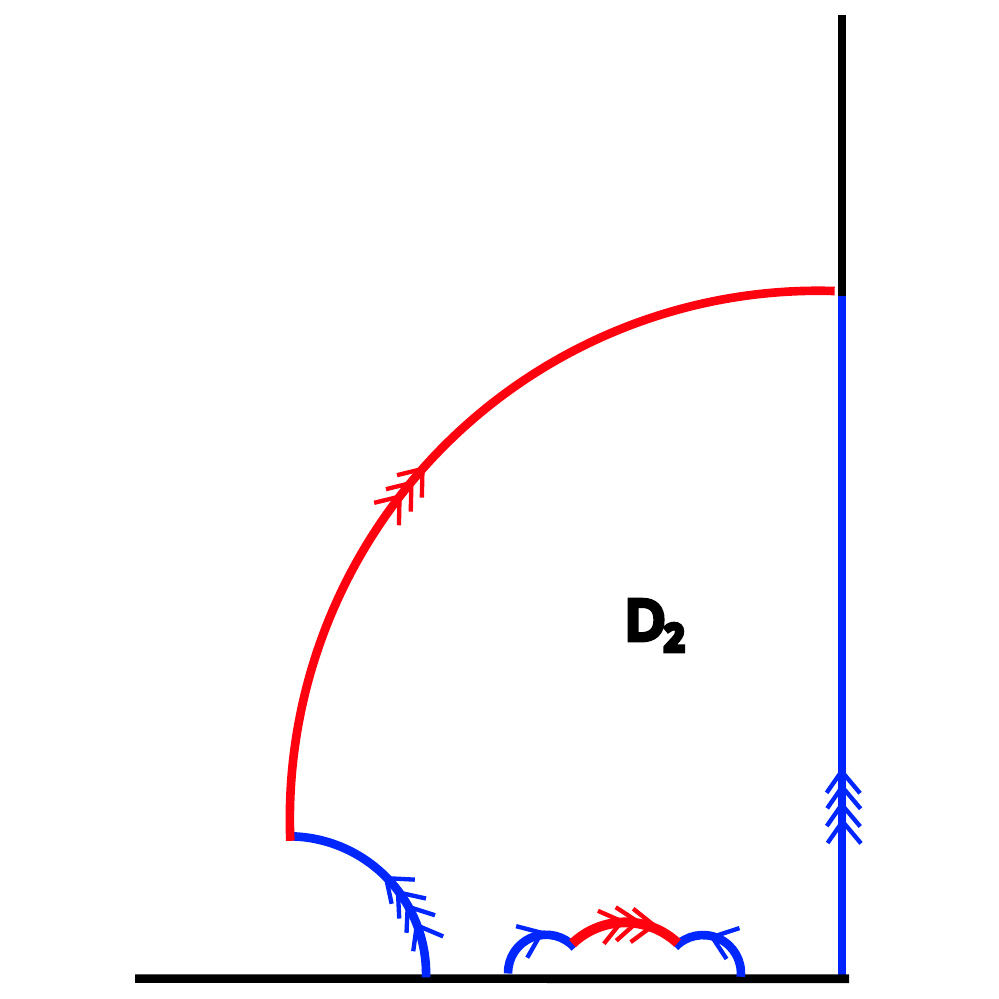}}

(a) \hspace{8.5cm} (b) 

\caption{{\bf (a)} As in Figure \ref{fig:halfplaneD1}(b), but showing also a geodesic arc (in red) that cuts $D_1$ in two symmetric halves, and the image of this arc in the shaded image of~$D_1$.
{\bf (b)} A new fundamental domain $D_2$, combining half of $D_1$ and half of the shaded image of $D_1$. The upper triple-arrowed boundary is mapped to the lower triple-arrowed boundary by ${\hat A}$. The left single-arrowed boundary is mapped to the right single-arrowed boundary by ${\hat A}{\hat B}{\hat A}^{-1}$.}
\label{fig:halfplaneD1-toD2}
\end{figure}

\begin{figure}[t]
\centering
\includegraphics[width=10 cm]{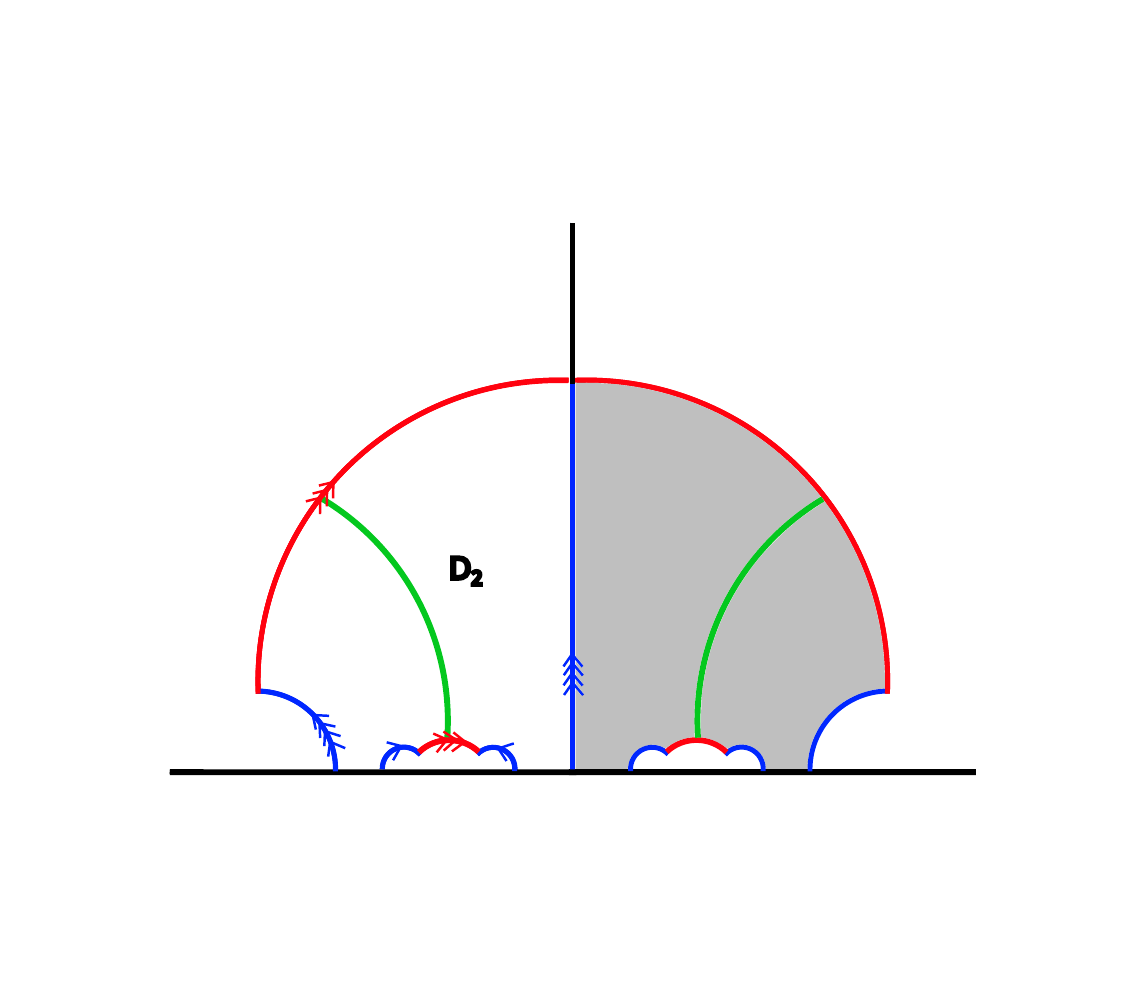}
\raisebox{6ex}[0pt][0pt]{\includegraphics[width=7.5 cm]{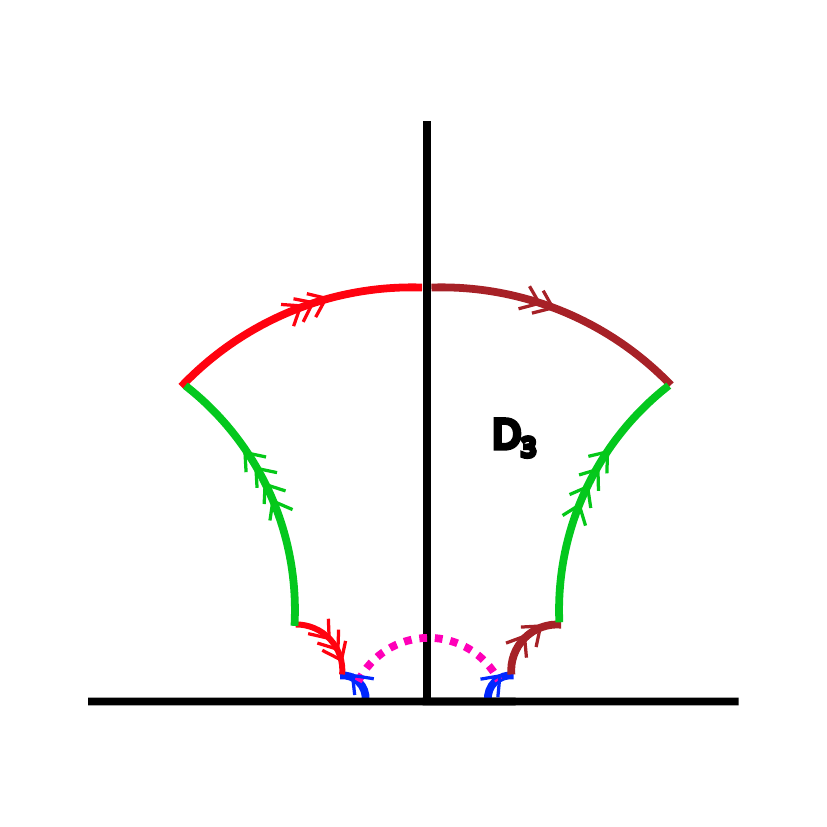}}

(a) \hspace{8cm} (b) 

\caption{{\bf (a)} $D_2$ together with its image under ${\hat B}$, shown shaded. Also shown is a geodesic arc (in green) that cuts $D_2$ in two symmetric halves, and the image of this arc in the shaded image of~$D_2$.
{\bf (b)} A new fundamental domain $D_3$, combining half of $D_2$ and half of the shaded image of~$D_2$. The left quadruple-arrowed boundary is mapped to the right quadruple-arrowed boundary by~${\hat B}$. The left single-arrowed boundary is mapped to the right single-arrowed boundary by 
${\hat B}{\hat A}{\hat B}^{-1}{\hat A}^{-1}$, 
whose fixed points are $w=\pm1$. The dashed curve is the geodesic circle joining $w=\pm1$ and forming the horizon.}
\label{fig:halfplaneD2-toD3}
\end{figure}

\section{Topology\label{sec:topology}}

In this appendix we show that the quotient surface of the Poincar\'e disc is a punctured torus, and its fundamental group is the free group generated by $A/\{\pm1\}$ and $B/\{\pm1\}$, were $A$ and $B$ are given in~\eqref{eq:poincareAB}. 

We start by representing the punctured torus as shown in Figure \ref{fig:punctured-torus-fd-basic-pluswithp}(a), as a punctured rectangle with the opposing sides pairwise identified as shown. We then perform a sequence of cut-and paste deformations as shown in Figures \ref{fig:punctured-torus-fd-basic-pluswithp}(b)--\ref{fig:punctured-torus-fd-deform-to-halfplane+perspective}(a). The fundamental domain in Figure \ref{fig:punctured-torus-fd-deform-to-halfplane+perspective}(a) is topologically identical to that in Figure \ref{fig:poincarediskdomains}(b), and the identifications by $A$ and $B$ in Figure \ref{fig:poincarediskdomains}(b) correspond to the pair of identifications across the boundaries of the fundamental domain in Figure \ref{fig:punctured-torus-fd-basic-pluswithp}(a). A perspective picture of the fundamental domain of Figure \ref{fig:halfplaneD2-toD3}(b) glued across the identifications is shown in Figure~\ref{fig:punctured-torus-fd-deform-to-halfplane+perspective}(b). 

As the fundamental group of the punctured torus is the free group generated by the two identifications shown in Figure \ref{fig:punctured-torus-fd-basic-pluswithp}(a) 
\cite[Section 1.2]{hatcherbook}\cite{pucturedtorusvideo}, it follows that $A/\{\pm1\}$ and $B/\{\pm1\}$ provide a pair of free generators for the action of the fundamental group on the Poincar\'e disc.

\begin{figure}[p]
\centering
\raisebox{-1ex}[0pt][0pt]{\includegraphics[width=8.5 cm]{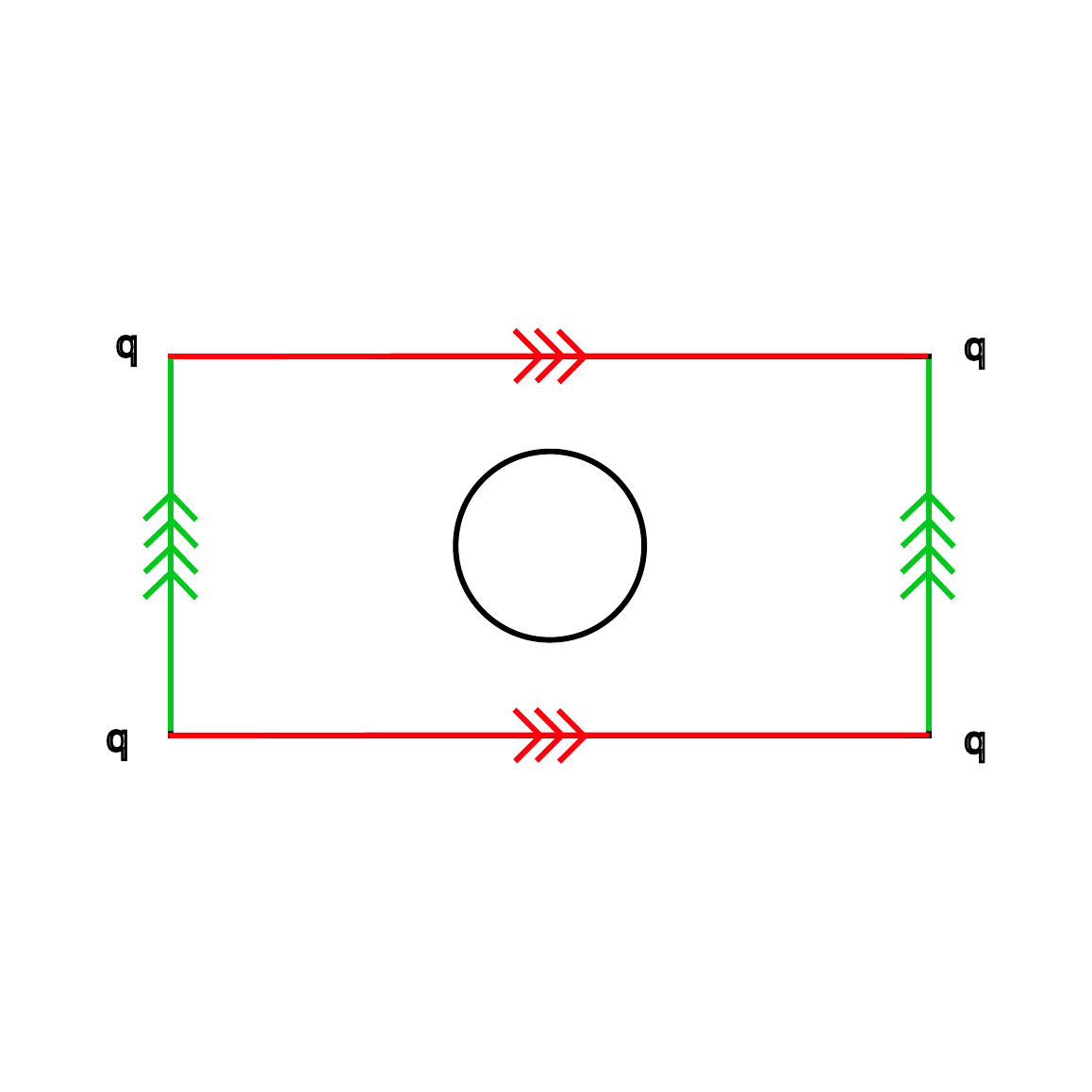}} 
% \hspace{1ex}
\includegraphics[width=8.0 cm]{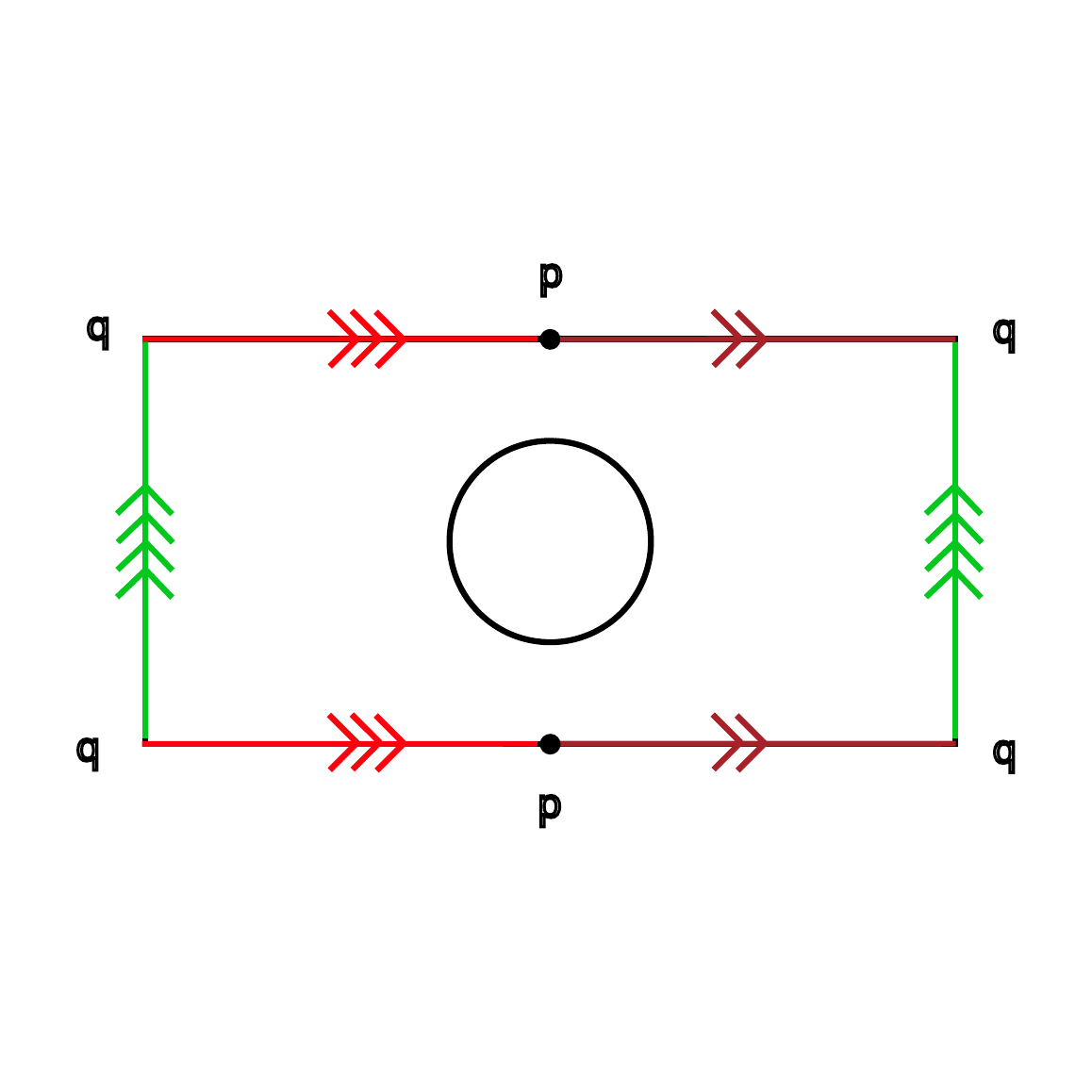}

(a) \hspace{7.5cm} (b) 

\caption{{\bf (a)} Fundamental domain of a punctured torus, 
realised as a rectangle with a hole in the middle. 
The arrows show the pairwise identifications of the sides.
All the four corner points, labelled $q$, are identified.
{\bf (b)} A pair of identified points labelled $p$ on the boundary are chosen. The double-arrowed boundary segments are identified, and the triple-arrowed boundary segments are identified.}
\label{fig:punctured-torus-fd-basic-pluswithp}
\end{figure}

\begin{figure}[h!]
\centering
\includegraphics[width=8.5 cm]{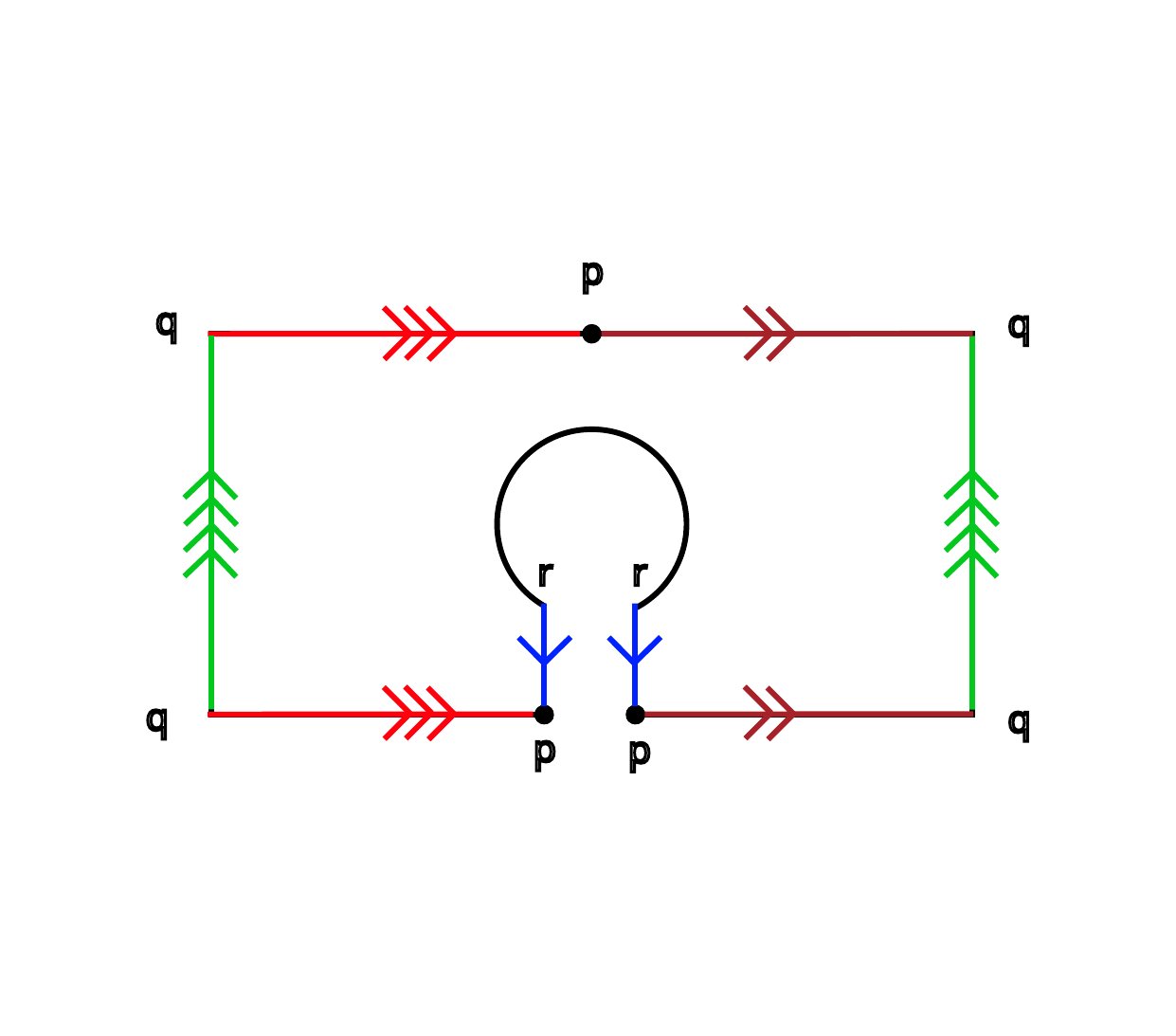} 
\raisebox{-5.5ex}[0pt][0pt]{\includegraphics[width=8.5 cm]{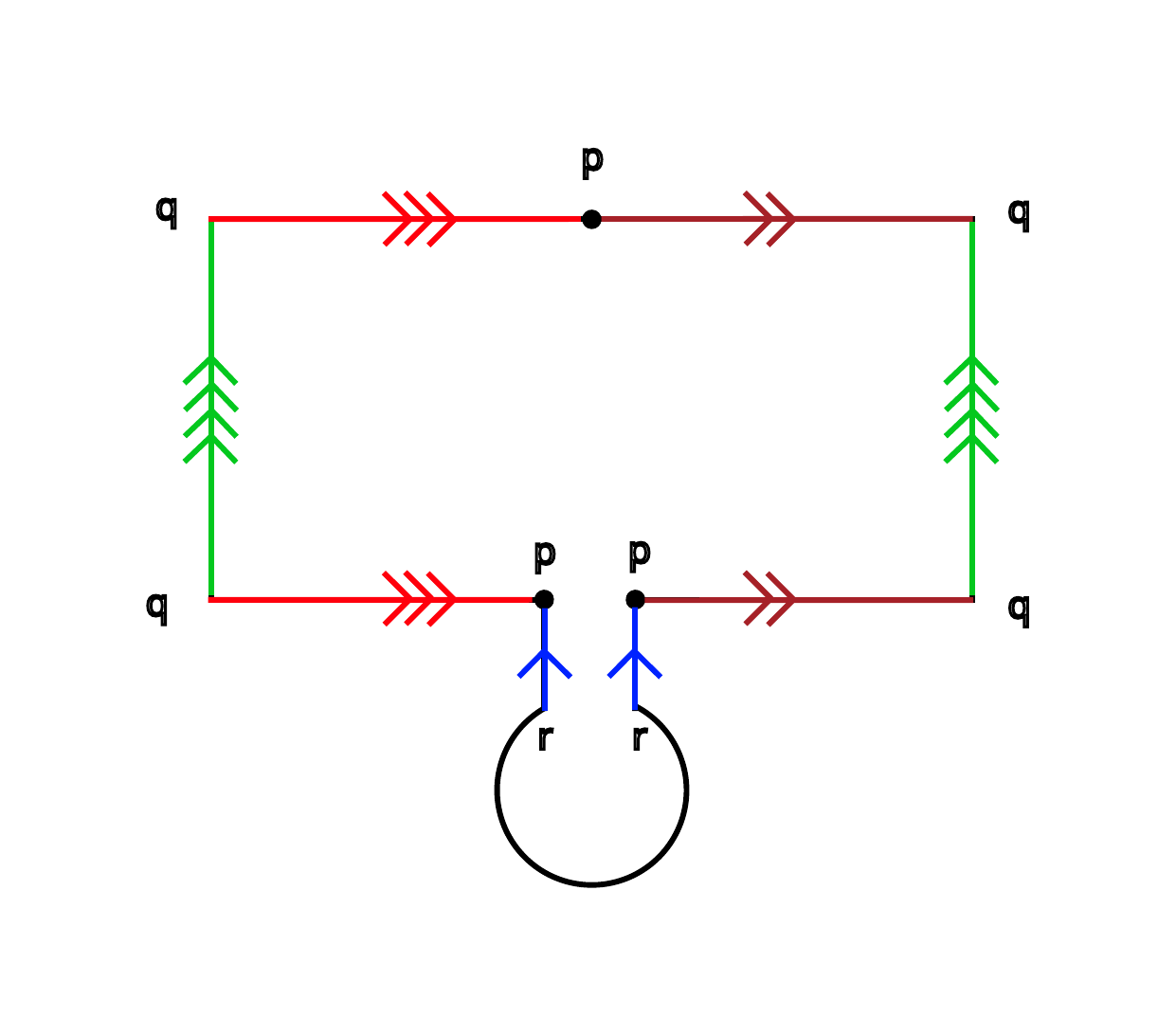}}

(a) \hspace{7.5cm} (b) 

\caption{{\bf (a)} A cut is introduced from one of the points $p$ to the puncture.
{\bf (b)} The surface is continuously deformed around the new cut.}
\label{fig:punctured-torus-fd-p-cut+deform}
\end{figure}

\begin{figure}[h!]
\centering
\raisebox{-10ex}[0pt][0pt]{\includegraphics[width=9.0 cm]{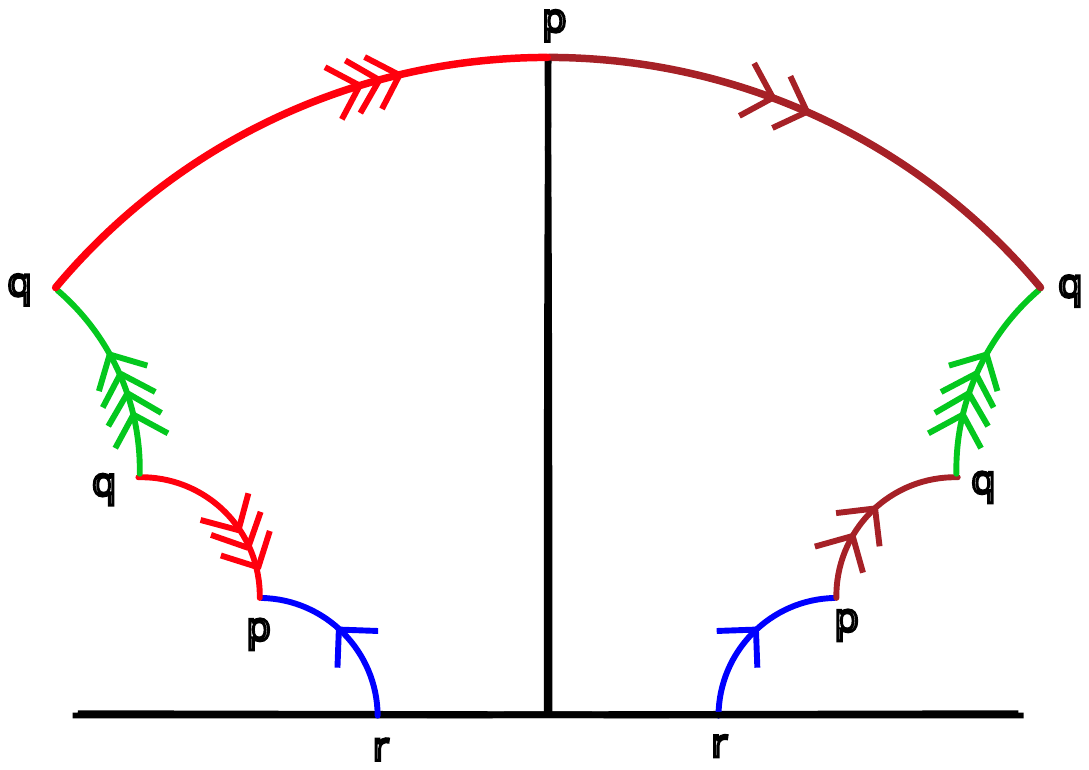}} 
\hspace{1ex}
\includegraphics[width=7.0 cm]{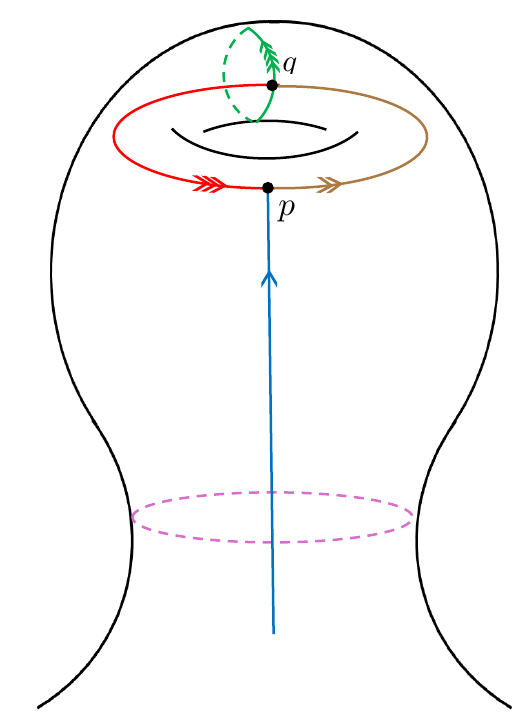}

(a) \hspace{9cm} (b) 

\caption{{\bf (a)} The straight boundary segments of Figure \ref{fig:punctured-torus-fd-p-cut+deform}(b) are deformed to curved boundary segments that match those in the upper half-plane fundamental domain in Figure~\ref{fig:halfplaneD2-toD3}(b).
{\bf (b)} A perspective picture in which the fundamental domain of panel (a) has been embedded in $\mathbb{R}^3$ and glued across the identified boundaries. The top side of the surface in panel (a) has become the inner side in panel~(b).}
\label{fig:punctured-torus-fd-deform-to-halfplane+perspective}
\end{figure}

\end{widetext}

\newpage 

$\phantom{xxx}$

\newpage 

$\phantom{xxx}$

\newpage 

\bibliography{ref}

\end{document}